\documentclass[5p, times,twocolumn]{elsarticle}

\usepackage{graphicx}
\usepackage{epsfig}
\usepackage{amssymb}
\usepackage{amsthm}
\usepackage{color}

\biboptions{sort, numbers}

\def\aj{AJ}%
\def\apj{ApJ}%
\def\apjl{ApJ}%
\def\aap{A\&A}%
\def\jcap{J. Cosmology Astropart. Phys.}%
\def\mnras{MNRAS}%
\def\prd{Phys.~Rev.~D}%
\def\nat{Nature}%

\newcommand{\eq}[1]{Eq.~(\ref{#1})}
\newcommand{\beq}{\begin{equation}}
\newcommand{\eeq}{\end{equation}}
\newcommand{\ud}{\mathrm{d}}
\newcommand{\lcdm}{{\ifmmode \Lambda{\rm CDM} \else $\Lambda{\rm CDM}$\fi}}
\newcommand{\Msol}{\rm{M}_\odot}
\newcommand{\Rsol}{\rm{R}_\odot}

\newcommand{\Ntot}{N_{\rm tot}}

\newcommand{\remove}[1]{}

\journal{Computer Physics Communications}

\begin{document}

\begin{frontmatter}



\title{{\sc clumpy}: a code for $\gamma$-ray signals from dark matter structures}

\author[label1]{Ald\'ee Charbonnier}
\author[label2,label3]{C\'eline Combet}
\ead{celine.combet@lpsc.in2p3.fr}
\author[label1,label2,label3,label4]{David Maurin}

\address[label1]{Laboratoire de Physique Nucl\'eaire et Hautes Energies,
  CNRS-IN2P3/Universit\'es Paris VI et Paris VII,
  4 place Jussieu, Tour 33, 75252 Paris Cedex 05, France}
\address[label2]{Dept. of Physics and Astronomy, University of
  Leicester, Leicester, LE1 7RH, UK}
\address[label3]{Laboratoire de Physique Subatomique et de Cosmologie,
CNRS/IN2P3/INPG/Universit\'e Joseph Fourier Grenoble 1,
53 avenue des Martyrs, 38026 Grenoble, France}
\address[label4]{Institut d'Astrophysique de Paris, UMR7095 CNRS,
Universit\'e Pierre et Marie Curie, 98 bis bd Arago, 75014 Paris, France}

\begin{abstract}
We present the first public code for semi-analytical calculation of the
$\gamma$-ray flux astrophysical $J$-factor from dark matter annihilation/decay 
in the Galaxy, including dark matter substructures. The core of the code is the calculation of
the line of sight integral of the dark matter density squared (for
annihilations) or density (for decaying dark matter). The code can be
used in three modes: i) to draw skymaps from the Galactic smooth component and/or the
substructure contributions, ii) to calculate the flux from a specific
halo (that is not the Galactic halo, e.g. dwarf spheroidal galaxies)
or iii) to perform simple statistical operations from a list of
allowed DM profiles for a given object. Extragalactic contributions and other tracers of DM
annihilation (e.g. positrons, antiprotons) will be included in a second
release.
\end{abstract}

\end{frontmatter}
%
%
{\bf PROGRAM SUMMARY}

\begin{small}
\noindent
{\em Program Title:} CLUMPY                                   \\
{\em Programming language:} C/C++                           \\
{\em Computer:} PC and Mac                                         \\
{\em Operating system:} UNIX(Linux), MacOS X                    \\
{\em RAM:} depends on the requested size of skymaps ($\sim$40 Mb for a $500 \times 500$ map)                                          
{\em Classification:} 1.1, 1.9               \\
{\em External routines/libraries:} CERN ROOT\footnote{{\tt http://root.cern.ch}}, 
Doxygen\footnote{\tt http://www.doxygen.org} (optional) \\
{\em Nature of problem:} Calculation of $\gamma$-ray signal from dark matter annihilation (resp. decay).
This involves a particle physics term and an astrophysical one. The focus here is on the latter.
\\
{\em Solution method:} Integration of the DM density squared (resp. density) along a line of sight. 
The code is optimised to deal with the DM density peaks 
encountered along the line of sight (DM substructures). 
A semi-analytical approach (calibrated on N-body simulations) is used for the 
spatial and mass distributions of the dark matter substructures in the Galaxy. 
   \\
{\em Restrictions:}
Some generic dark matter annihilation spectra are provided but are not included in the calculation so far as it is assumed
that the particle physics is independent of the astrophysics of the problem.  
   \\
{\em Running time:} this is highly dependent of the DM profiles considered, the requested precision $\epsilon$ and integration angle $\alpha_{\rm int}$:
\begin{itemize}
\item about 60 mn for a $5^\circ \times 5^\circ$ map towards the Galactic centre, with $\alpha_{\rm int}=0.01^\circ$, NFW dark matter profiles and $\epsilon=10^{-2}$; 
\item about 2h for the same set-up towards the anti-centre;
\item 0.1 to 10 DM models per second, depending on integration angle and DM profile. 
\end{itemize}

\end{small}


\section{Introduction \label{sec:intro}}

Despite the several astrophysical evidences pointing at the existence of
dark matter (flat rotation curves of galaxies, gravitational lensing,
``bullet cluster", etc.), its nature still evades us. This search has
become one of the major topic in both particle physics and  astrophysics
and  is tackled using either direct or indirect detection methods. For
the former, the hope is to directly witness the interaction of a dark
matter particle with a detector. The indirect approaches aim at measuring
the end products of dark matter annihilation/decay (e.g., $\gamma$-rays,
positrons, anti-protons).  The detection of $\gamma$-rays
was soon recognised to be a promising channel
\citep{1978ApJ...223.1015G,1978ApJ...223.1032S}. If such a signal is yet
to be measured by the existing $\gamma$-rays observatories (FERMI, HESS,
MAGIC, VERITAS), the prospect may significantly improve with the
forthcoming next-generation instruments, such as CTA
\citep{2010arXiv1008.3703C}. 

Having a modelling tool that can compute the expected $\gamma$-ray flux
from dark matter annihilation/decay in a wide range of astrophysical
configurations and making it available to the community  should prove
very useful. This is the motivation for developing the code presented in
this paper. This paper highlights {\sc clumpy}'s main features but a more
thorough description  can be found in the documentation coming with the source code.  The paper is
organised as follows. Section~\ref{sec:calculation} presents the
formulation of the  generic $\gamma$-ray signal calculation that is
performed by {\sc clumpy}, with special emphasis on  the contribution
from DM clumps. The code is described in sect.~\ref{sec:code}, which
presents the structure, parameters and functions that are at the core of
{\sc clumpy}. Section~\ref{sec:examples} deals with two examples of the
science that can be performed with {\sc clumpy}. We conclude and discuss
the future developments of the code in sect.~\ref{sec:conclusion}.

\section{Calculating the $\gamma$-ray flux from dark matter annihilation/decay\label{sec:calculation}}

The $\gamma$-ray flux $\ud \Phi_{\gamma}/\ud E_{\gamma}$ from dark matter
annihilating/decaying particles is expressed as the product of a particle physics
term by an astrophysical contribution.  For a given experiment with a
spatial resolution $\alpha_{\rm int}$ corresponding to an integration
solid angle $\Delta \Omega=2\pi\,(1-\cos\,\alpha_{\rm int})$, at  energy
$E_{\gamma}$ and pointing in the direction $(\psi,\theta)$, the flux is
written as:
\beq
\frac{\ud \Phi_{\gamma}}{\ud E_{\gamma}}(E_{\gamma},\psi,\theta,\Delta
\Omega)=\frac{\ud \Phi_{\gamma}^{PP}}{\ud
  E_{\gamma}}(E_{\gamma})\times J(\psi,\theta,\Delta \Omega) \, \textrm{.}
\label{eq:flux-general}
\eeq

\subsection{Annihilation and decay: general statements}

\subsubsection{Annihilation: the particle physics term \label{subsec:pp-term}}

It can be generically expressed as
\beq
  \frac{\ud \Phi_{\gamma}^{PP}}{\ud E_{\gamma}}(E_{\gamma})
  \equiv \frac{1}{4\pi}\, \frac{\langle \sigma_{\rm ann}v \rangle}{2m_{\chi}^{2}}
  \cdot \sum_{f}\frac{dN^{f}_{\gamma}}{dE_{\gamma}}\, B_{f}\,,
  \label{eq:term-pp}
\eeq
where $m_\chi$ is the mass of the DM particle, $\sigma_{\rm ann}$ is the
annihilation cross section, $\langle\sigma_{\rm ann}v\rangle$ the
annihilation rate averaged over the DM velocity distribution, $B_f$ is the
branching ratio into the final state $f$ and $dN^{f}_{\gamma}/dE_{\gamma}$
is the photon yield per annihilation.  A widely accepted value for
$\langle\sigma_{\rm ann}v\rangle$ estimated from present day thermal 
relic density is $\langle\sigma_{\rm ann}v\rangle = 3\cdot
10^{-26}$~cm$^{3}$s$^{-1}$ \citep{1998APh.....9..137B}\footnote{Note
that a velocity dependent cross section should be included in a future
release of the code in order to account for the Sommerfeld effect in the
calculation. This effect results in much larger cross sections for some
resonant neutralino masses and therefore boost the signal for these DM
candidates.}. The photon
spectrum is dependent on the annihilation channels. We restrict ourselves
to the popular \emph{Minimal Supersymmetric Standard Model} (MSSM). In
this framework, the neutralino is the lightest stable supersymmetric
particle and is the favoured DM candidate.

Neutralino annihilation can produce $\gamma$-rays in three ways: i) DM
annihilates directly into two photons or $Z^0\gamma$, giving rise to
monochromatic lines, ii) or it can annihilate into primary products, the
hadronization and decay of which produce a $\gamma$-ray continuum, iii)
finally, if charged particles are produced, internal bremsstrahlung (IB)
can also contribute to the continuum. The lines are very model-dependent
and given the uncertainties on the particle physics models, we did not
include any in {\sc clumpy}. The user is however free to implement their own
function if need be. There exists in the literature several
parametrisations for the $\gamma$-ray continuum and we
have implemented that of Bergstr\"om et al. \citep{1998APh.....9..137B}, and
Tasitsiomi and Olinto \citep{2002PhRvD..66h3006T}. The IB contributions is also very model
dependent, and we provide one benchmark model from
Bringmann et al. \citep{2008JHEP...01..049B}. We refer the reader to
\S\ref{subsubsec:libpp} ({\tt spectra.h}) for more details
on these spectra.

\subsubsection{Annihilation: the astrophysical contribution \label{subsec:term-astro}}

The $\gamma$-rays being produced from the annihilation of pairs of dark matter
particles, the $\gamma$-ray flux is proportional to the DM density
squared. The second term of Eq.~(\ref{eq:flux-general}), termed
\emph{astrophysical} contribution, corresponds to the integration of the
density squared along the line of sight ($l, \psi,\theta$), in the
observational cone of solid angle $\Delta\Omega$ (see \ref{app:geom}
and {\tt geometry.h} for definitions and the geometry used)
\beq
J(\psi,\theta, \Delta \Omega) = \int_{0}^{\Delta \Omega}\int_{\rm{l.o.s}} \rho^{2}( l(\psi,\theta)) \;
 \;\ud l \, \ud \Omega \,\;,
\label{eq:term-astro}
\eeq
where $\rho(l(\psi,\theta))$ is the local DM density at distance $l$ from
Earth in the direction $\psi,\theta$ (longitude and latitude in Galactic
coordinates).

This integral is the core of the {\sc clumpy} code and
can be rather complicated to evaluate. The difficulty comes partially from
the  steepness of some DM profiles considered but also, and mainly, from
the existence of DM clumps, the contribution of which must be added to the
smooth Galactic DM background. Both components are dealt with hereafter.

\subsubsection{Decaying dark matter}
$\gamma$-rays can also
be produced from the decay of DM. Formally, the flux is still given by \eq{eq:flux-general}
but the astrophysical and particle physics terms take different forms:
\begin{itemize}
\item The particle physics term now corresponds to the decay spectrum and no parametrisation
of that term has been included in the code so far. The parametrisation 
from Bertone et al. \citep{2007JCAP...11..003B} could be used.
\item Because we are now considering a decay reaction, the astrophysics term is the integration 
of the density (rather that density squared as in \eq{eq:term-astro}) along the line of sight, namely
\beq
J(\psi,\theta, \Delta \Omega) = \int_{0}^{\Delta \Omega}\int_{\rm{l.o.s}} \rho( l(\psi,\theta)) \;
 \;\ud l \, \ud \Omega \,\;.
\label{eq:term-astro2}
\eeq
This quantity can be straightforwardly computed in {\tt CLUMPY} by setting the flag parameter 
{\tt gDM\_IS\_ANNIHIL\_OR\_DECAY} to {\tt false} (see table~\ref{tab:param}). 
\end{itemize}

\subsection{Dark matter distribution}
To date, there is no consensus to what the Galactic
DM profile, $\rho(r)$, should be. There is some dynamical
evidence that the Galactic DM halo might be triaxial 
\citep{2009ApJ...703L..67L} and numerical simulations, 
such as the Aquarius \citep{2008MNRAS.391.1685S}  or the Via Lactea runs
\citep{2008Natur.454..735D} also find non-spherical halos. When included,
the baryons tend however to make the halos more spherical \citep{2004ApJ...611L..73K}.
In this first version of {\sc clumpy} we simplify the problem  (as often done)
using spherically symmetric DM halos, but allowing for tri-axiality is being
considered for future releases. 
The total averaged density
profiles of the DM halos  measured in these simulations are generally fitted by
a form of  the Zhao profile \citep{1996MNRAS.278..488Z} :
\beq
\rho_{\rm tot}(r)=\frac{\rho_s}{\left(r/r_s\right)^\gamma\,\left[1+(r/r_s)^\alpha\right]^{(\beta-\gamma)/\alpha}}\;,
\label{eq:zhao}
\eeq
where $\gamma$, $\alpha$ and $\beta$ are respectively the inner, transition
and outer slope of the profile, $\rho_s$ and $r_s$ a scale density and
radius.  The Navarro et al. \citep{1997ApJ...490..493N}, Moore et al. \citep{1998ApJ...499L...5M},
and Diemand et al. \citep{2004MNRAS.353..624D} profiles fall in this category. Some other
profiles having a log-varying inner slope \citep{2004MNRAS.349.1039N,2006AJ....132.2685M} are also included
in {\sc clumpy}. The reader is referred to \S\ref{subsubsec:libprofiles} 
({\tt profiles.h}) for a detailed description. 

Some of these DM profiles are steep enough in the inner regions to lead to a
singularity of the $\gamma$-ray luminosity at the centre of the halo.
However, a cut-off radius $r_{\rm{cut}}$ naturally appears, within  which
the DM density saturates due to the balance between the annihilation rate
and  the gravitational infalling rate of DM particles. The saturation
density reads \citep{1992PhLB..294..221B} 
\beq
\rho_{\rm sat}= 3.10^{18}
     \left(\frac{m_\chi}{100~\rm GeV}\right) \times
     \left( \frac{10^{-26} {\rm cm}^3~{\rm s}^{-1}}
	  {\langle \sigma v\rangle}\right)
   \Msol~{\rm kpc}^{-3}.
  \label{eq:rho_sat} 
\eeq
Plugged in any profile parametrisation, this saturation density defines
the cut-off radius below which the annihilation rate is constant. This is a
very crude description, but this is not important as this cut-off matters
only for the steepest profiles ($\gamma\ge1.5$) which are disfavoured by
current numerical simulations.

 The hierarchical scenario of galaxy formation in a $\Lambda$-CDM cosmology is
characterised by a high degree of clumpiness of the DM
distribution (supported by N-body
simulations) so that the total averaged density computed from these
simulations corresponds to
\beq
\rho_{\rm tot} (r) = \rho_{\rm sm}(r) + \langle \rho_{\rm
  subs}(r)\rangle
\label{eq:rhotot}
\eeq
where $ \rho_{\rm sm}(r)$ is the ``true'' smooth component and $\langle \rho_{\rm
  sub}(r)\rangle$  represents the average density from the substructures. These two
components are discussed below.

\subsubsection{The clumps \label{subsubsec:clumps}}
Whether substructures in subhalos are scaled-down versions of substructure in main halos remains an open 
question \citep{2008MNRAS.391.1685S}. In this release of {\sc clumpy}, we only consider one level of substructure within 
the halo under scrutiny (Galactic halo or individual halo such as a dwarf spheroidal galaxy), and the properties of these sub-halos 
can be independently chosen from that of the parent halo (see below).

\paragraph{Clump individual properties}
In the code, all clumps have the same inner DM distribution, which can
be any of the profiles discussed in \S\ref{subsubsec:libprofiles}.
The mass of the clump generally suffices to determine all its
properties, i.e., its size
$R_{\rm vir}^{\rm cl}$, and once an inner density profile is chosen,  its scale
radius $r_s$ and scale density $\rho_s$. This is done using the so-called
\emph{concentration parameter}  $c_{\rm vir}$. Parametrisations of the
mass-concentration relation have  been established from numerical simulations for
isolated field halos \citep{2001MNRAS.321..559B,2001ApJ...554..114E} but the
concentration  generally presents a significant scatter
\citep{2002ApJ...568...52W}. These relations have been implemented in the code,
regardless of that scatter or of the environment and formation history of the
halos that can also affect the value of the concentration. We refer the reader to
\S\ref{subsubsec:libclump} ({\tt clumps.h}) and the Doxygen documentation for more
details.

\paragraph{Spatial and mass distribution of the clumps} Following
Lavalle et al. \citep{2008A&A...479..427L}, spherical symmetry
is assumed and the two distributions are assumed 
independent\footnote{Keep in mind that such an assumption fails near the
Galactic centre, where small clumps  are expected to be disrupted by tidal
forces. In this direction, the signal from the smooth
contribution dominates anyway.}. If $\Ntot$ is the total number of
clumps in a DM host halo, then the overall distribution of clumps is written
as:
\beq
\frac{\ud^2N}{\ud V \ud M} = \Ntot\, \frac{\ud{\cal P}_V(r)}{\ud V}\,\frac{\ud {\cal P}_M(M)}{\ud M} \ \textrm{,}
\label{eq:distribution-clumps}
\eeq
where the spatial and mass distribution are probabilities,
respectively normalised as:
\beq
\int_{\rm V} \frac{\ud{\cal P}_V(r)}{\ud V}\, \ud V = 1 \qquad \textrm{and} \qquad \int_{M_{\rm min}}^{M_{\rm max}} \frac{\ud {\cal P}_M(M)}{\ud M}\, \ud M = 1 \, .
\label{eq:normalisation-clumps}
\eeq
Analysis of N-body simulations have shown the mass distribution to vary as a
simple power law
\beq
\frac{\ud {\cal P}_M(M)}{\ud M} \propto M^{-\alpha_M}\,\; \textrm{with } \alpha_M\approx 1.9\,.
\label{eq:clump_mass_function}
\eeq
Given the mass and spatial distribution, the total number of clumps
$N_{\rm tot}$ can be determined in two ways. If the mass fraction $f$ of
clumps and the mass of the host $M_{\rm tot}$ halo are known,
\[
N_{\rm tot}=\frac{fM_{\rm tot}}{\langle M_{\rm 1cl}\rangle}\;,
\]
with $\langle M_{\rm 1cl}\rangle$ the average mass of one clump. An other
approach is to know the number of clumps $N_0$ in a given mass interval
$[M_1,M_2]$ and to calculate the total number of clumps as
\[
N_{\rm tot}=\frac{N_0 }{ \int_{M_1}^{M_2} \frac{\ud {\cal P}_M(M)}{\ud M}}\;.
\]
{\sc Clumpy} uses the first approach when dealing with substructures
in individual host halos (that are not the Milky Way) and the second
one for the Galactic halo clumps (numerical simulations have found 
about 100 clumps with masses above $10^8$~M$_\odot$ in Milky Way-like
halos).
{\sc clumpy} allows to select
any of the profiles given in \S\ref{subsubsec:libprofiles} ({\tt
profiles.h}) for the clump spatial distribution, independently of the
choice made for the total density profile.

\subsubsection{The smooth component \label{subsubsec:smooth}}
From the spatial distribution ${\rm d}{\cal P}_V(r)/\ud V$ one can
define the average clump density $\langle \rho_{\rm cl} \rangle$ used
in \eq{eq:rhotot} as:
\beq
\langle \rho_{\rm subs}(r) \rangle = f M_{\rm tot} \frac{\ud{\cal P}_V(r)}{\ud V}\;.
\label{eq:rhocl}
\eeq
For a given total density profile and clump spatial distribution,
\eq{eq:rhotot} then gives the smooth density profile as 
\beq
\rho_{\rm sm}(r)=\rho_{\rm tot}(r) -\langle \rho_{\rm subs}(r)
\rangle\;.
\label{eq:rhosm}
\eeq
Using this approach, rather than providing independently a smooth and
a clump profile, ensures (by construction) that the total density
profiles follows the one measured in N-body simulations. This has been
discussed in Pieri et al. \citep{2011PhRvD..83b3518P}.

\subsection{Calculating the annihilation J-factor\label{subsec:calc_J}}
The astrophysical contribution to the annihilation flux is formally given by
\eq{eq:term-astro}. The J-factor is more explicitly written as:
\beq
J = \int_0^{\Delta\Omega}\int_{l_{\rm min}}^{l_{\rm max}} \frac{1}{l^2}\left( \rho_{\rm sm} + \sum_i \rho_{\rm
    cl}^i\right)^2 l^2 dl d\Omega\;.
\eeq
where $\rho_{\rm sm }$ is given by \eq{eq:rhosm} and the second term
corresponds to the sum of the inner densities squared of all clumps $i$
contained with the volume element. The latter should not be confused
with $\langle \rho_{\rm subs}(r) \rangle$.
Three terms arise from this equation:
\beq
J_{\rm sm}\equiv \int_0^{\Delta\Omega}\int_{l_{\rm min}}^{l_{\rm max}}
\rho_{\rm sm}^2 dl d\Omega\;,
\label{eq:gal_Jsm}
\eeq
\beq
J_{\rm subs}\equiv \int_0^{\Delta\Omega}\int_{l_{\rm min}}^{l_{\rm max}} \left(\sum_i \rho_{\rm
    cl}^i\right)^2 dl d\Omega\;,
\eeq
and the cross product
\beq
J_{\rm cross-prod}\equiv 2\int_0^{\Delta\Omega}\int_{l_{\rm min}}^{l_{\rm max}}
\rho_{\rm sm}\sum_i \rho_{\rm
    cl}^i dl d\Omega\;.
\eeq

\subsection{Statistical considerations}

Taking $\alpha_M=1.9$ as a canonical value for the slope of the mass
distribution, with 100 clumps above $10^8$~M$_\odot$ as determined from
N-body simulations, and a minimal theoretical mass of $10^{-6}$~M$_\odot$
for the smallest ones, the total number of clumps in the  Milky Way halo
(i.e., for $\Delta\Omega=4\pi$) is estimated to be $\sim 10^{14}$, which is
computationally prohibitive. Even if we are interested in a
$5^\circ\times5^\circ$ skymap only (i.e., $\Delta\Omega\sim 6\times
10^{-3}$), the number of clumps is $\sim 5\times 10^{10}$. This is too
large a number to realise a statistical drawing according to the $\ud
{\cal P}_V/\ud V$ distribution and then numerically integrate the
$J$ component for each of these clumps. However, their high
number means that a statistical average can be safely performed, 
as long as the variance is small.
The {\sc clumpy} code computes the $\gamma$-ray flux from DM clumps
by either randomly drawing clumps given user-defined spatial and mass
distributions and inner DM profiles, or by calculating the mean contribution
whenever the variance of this contribution is smaller than some user-defined
value.  

\subsubsection{Averaged quantities from the continuum limit \label{subsubsec:average}}
Given the clump mass and spatial distributions given above, it is
possible to define average values for the primary random variables of the
problem (a clump mass and position) and the one deriving from them.
The mean mass and the mean luminosity are respectively defined by
\beq
\langle M\rangle = \int_{M_{\rm min}}^{M_{\rm max}} M\frac{\ud {\cal P}_M}{\ud M} {\ud M}\;,
\label{eq:gal_meanMcl} 
\eeq
\beq
\langle{\cal L}\rangle = \int_{M_{\rm min}}^{M_{\rm max}} {\cal L}(M)\frac{\ud {\cal P}_M}{\ud M} {\ud M}\;.
\label{eq:gal_meanLcl} 
\eeq
where ${\cal L}(M)$ is the intrinsic luminosity of a clump given by
\beq
{\cal L}(M)\equiv\int_{\rm V_{\rm cl}} \left(\rho_{\rm cl}\right)^2 \ud V\;.
\label{eq:luminosity}
\eeq
The average
of the distance to any power $n$ is written as
\beq
\langle l^n \rangle = \int_0^{\Delta\Omega} \int_{l_{\rm min}}^{l_{\rm
    max}} l^n \frac{\ud {\cal P}_V}{\ud V} l^2 dl d\Omega\;.
\label{eq:mean_dist}
\eeq
Combining these,  it can be shown that the average contribution from
the clumps is expressed as
\beq
\langle J_{\rm subs}\rangle = \Ntot \int_0^{\Delta\Omega} \int_{l_{\rm min}}^{l_{\rm max}}
\frac{{\rm d}{\cal P}_V}{\ud V} dld\Omega \;\int_{M_{\rm min}}^{M_{\rm max}} {\cal L}(M)\frac{\ud {\cal P}_M}{\ud M} {\ud M}\;,
\label{eq:gal_meanJcl} 
\eeq
In the same spirit, in the continuum limit the cross-product becomes
\beq
\langle J_{\rm cross-prod}\rangle=2
\int_0^{\Delta\Omega}\int_{l_{\rm min}}^{l_{\rm max}} \rho_{\rm
  sm}\langle \rho_{\rm subs}\rangle dld\Omega\;.
\label{eq:cross_prod}
\eeq
The importance of the cross product can easily be identified when
considering that the clump spatial distribution follows the total
average density profile. In that case, $\langle \rho_{\rm subs}(r)\rangle
= f\rho_{\rm tot}(r)$ and $\rho_{\rm sm}(r)
= (1-f)\rho_{\rm tot}(r)$, leading directly to $J_{\rm cross-prod}/J_{\rm
  tot} = 2f(1-f)$.
For a typical clump mass fraction $f=10\%$, this cross product amounts
to $18\%$ of the reference J-factor $J_{\rm tot}\equiv\int \rho_{\rm
  tot}^2 dld\Omega$ . This shows that when an averaged clump description is
used to integrate the signal along the line of sight, this cross
product is not negligible and should be taken into account. In
\ref{app:crossprod}, it is however shown that on an individual basis,
the cross product of one clump in an underlying smooth Galactic component
can be safely discarded. This will apply to the clumps that
need to be statistical drawn in our model (see below).

\subsubsection{Variance of the clump contribution $\sigma^2_{\rm clumps}$}

Along a given line of sight and for a given integration angle $\alpha_{\rm
int}$, the average flux between $l_{\rm min}$ and $l_{\rm max}$ given by
\eq{eq:gal_meanJcl} can also be expressed as
\beq
\langle J_{\rm subs}\rangle=\langle N_{\rm cl}\rangle \langle J_{\rm 1\,cl}\rangle\;,
\eeq
where $\langle N_{\rm cl}\rangle$ is the average number of clumps in this volume (given
the distance interval and mass range considered), and
$\langle J_{\rm 1\;cl}\rangle$ the average flux of one clump (see also
\citep{2007A&A...462..827L}). Using the point-like approximation, a
clump of mass $M$ at a distance $l$ has a flux
\beq
J_{\rm 1\,cl}=\frac{{\cal L}(M)}{l^2}\;,
\eeq
so that, recalling that the mass of the clump and its location are independent variables, 
\beq
\langle J_{\rm 1\,cl}\rangle=\langle{\cal L}\rangle \left<\frac{1}{ l^2}\right>\,.
\eeq
The variance on the flux of a single clump is then defined as
\beq
\sigma^2_{1\,\rm cl}= \langle J_{\rm 1\,cl}^2\rangle- \langle J_{\rm
  1\,cl}\rangle^2=\langle {\cal L}^2\rangle \left\langle\frac{1}{l^4}\right\rangle - \langle J_{\rm 1\,cl}\rangle^2\;.
\label{eq:var_J}
\eeq
The variance of a population of $\langle N_{\rm cl}\rangle $ clumps follows as
\beq
\sigma^2_{\rm cl}=\langle N_{\rm cl}\rangle\,\sigma^2_{1\,\rm cl}\,.
\eeq

\subsubsection{Criterion for using the averaged description \label{subsubsec:l_crit}}
For a given integration domain, there
is a threshold mass above which the clumps are not numerous enough
to be described by an averaged description. Conversely, for a given
mass decade there is a critical distance below which the averaged
description fails and where clumps should be statistically drawn. 

The averaged description can be safely used, as long as the relative
error $RE$ made on the clump flux with respect to the \emph{total} flux
(i.e. including the smooth contribution)
\beq
RE_{J_{\rm clumps}}=\frac{\sqrt{N_{\rm cl}}\;\sigma_{1\,\rm
    cl}}{N_{\rm cl}\langle J_{\rm 1\,cl}\rangle+J_{\rm smooth}}
\label{eq:RSE}
\eeq
is smaller than some user-defined prescription.

  \paragraph{Critical distance $l_{\rm crit}$ for the clumps in the Galaxy}

\begin{figure}[!t]
\begin{center}
\includegraphics[clip=,width=\columnwidth]{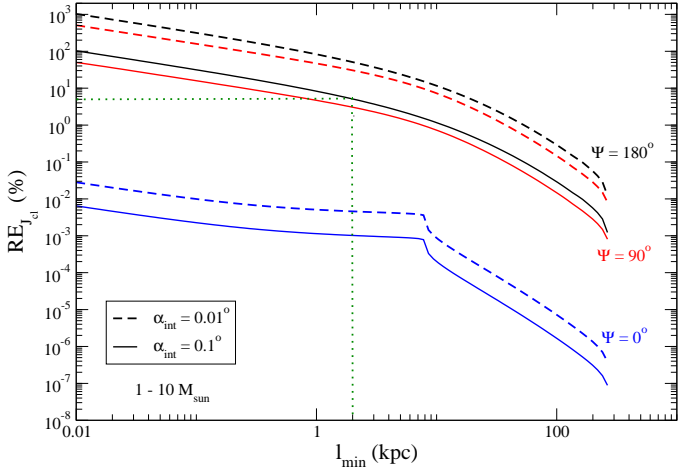}
\caption{Relative error between the \emph{average} and \emph{true} $J$
value from clumps when the signal is integrated between $l_{\rm
min}$ and $l_{\rm max}$, according to \eq{eq:RSE}, for $M_{\rm
cl}\in[1,10]$~M$_\odot$. The signal is integrated over 
$\alpha_{\rm int}=0.1^\circ$ (solid line) or $ \alpha_{\rm int}=0.01^\circ$
(dashed line). Three directions in the Galactic plane are considered:
towards the Galactic centre (blue), towards the  anti-centre (black) and
towards the Galactic East (red). For the latter the green dotted line
illustrates how to get the critical distance (here 2~kpc) below which
clumps (in this mass range) must be drawn if the user requires an accuracy
of 5\%. All distributions (smooth, clump distribution and inner DM clump profile) are taken as NFW profiles.\label{fig:RE}}
\end{center}
\vspace{-0.5cm}
\end{figure}

The RE is plotted in Fig.~\ref{fig:RE} as a function of the
lower limit of the integration $l_{\rm min}$. As can be seen, for any
required precision (e.g., 5\% in  Fig.~\ref{fig:RE}, tagged by the dotted
green line), one can infer the corresponding distance $l_{\rm crit}$ below
which the relative error becomes larger than the request.

\begin{figure*}[!t]
\begin{center}
\includegraphics[clip=,width=0.95\textwidth]{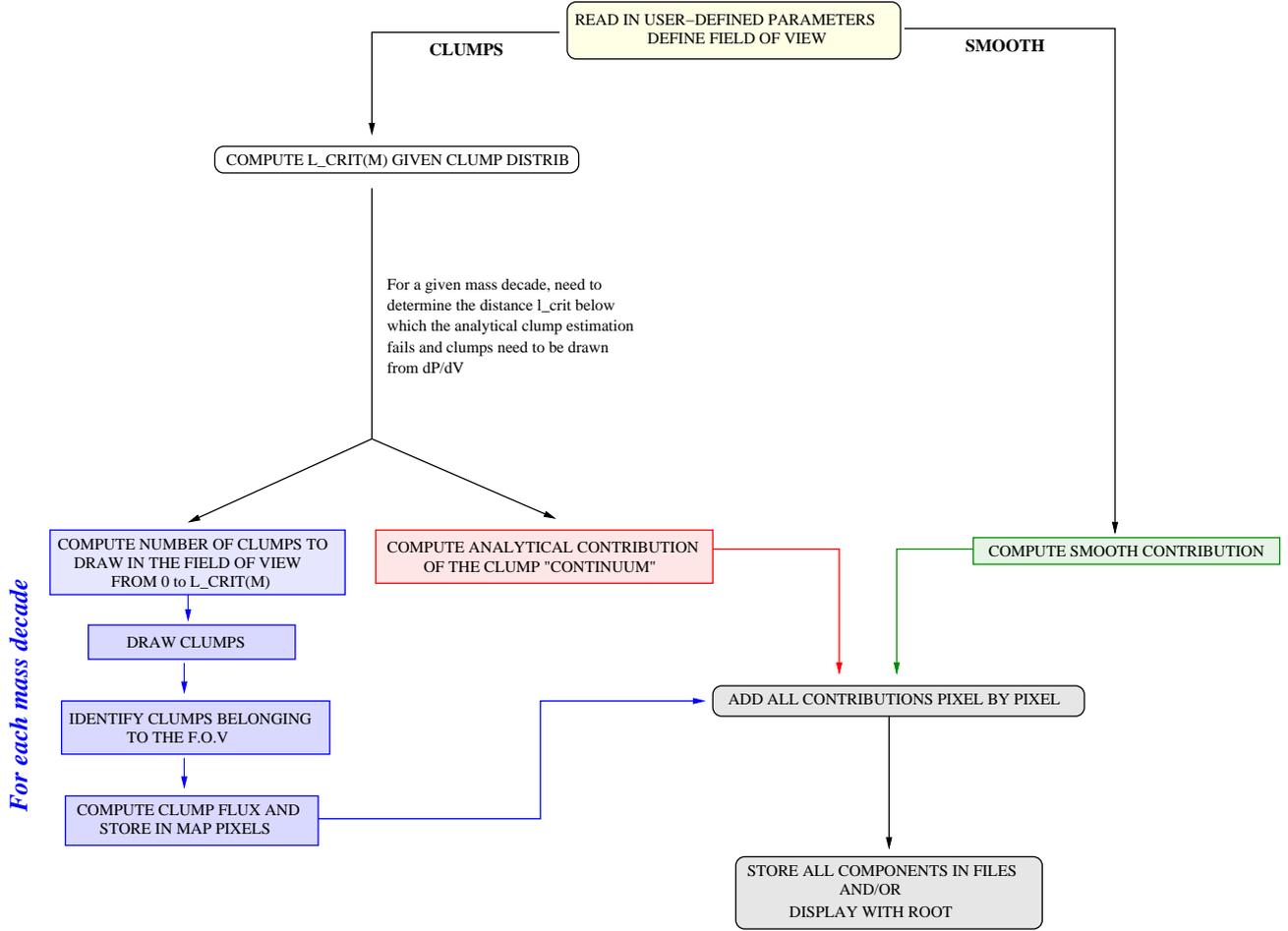}
\caption{Flow chart of the different steps leading to the creation of a
$\gamma$-ray skymap.}
\label{fig:flowchart}
\end{center}
\vspace{-0.5cm}
\end{figure*}

In practise, it is convenient to perform the calculation in terms of mass
decades: for each decade, we find (by dichotomy) the critical distance
below which clumps must be drawn\footnote{This mass decade corresponds to
a typical clump radius $R_{\rm cl}$ and we also check that the
point-like hypothesis used to derived the continuum limit is appropriate ($l_{\rm crit}\gg R_{\rm cl}$).}.
A lower limit of $10^{-3}$~kpc is set on
$l_{\rm crit}$ to avoid divergences for a clump sitting at the observer's
location. An example for the final number of clumps
to draw below $l_{\rm crit}$ is provided in Tab.~\ref{tab:l_crit}.
\begin{table}
\begin{center}
\caption{Example of $l_{\rm crit}$ and the associate number of
clumps to draw toward the anti-centre (criterion $RE_{J_{\rm clumps}}\!\!\!\!\!\!<5\%$)
for a $2^\circ\!\!\times\!\!2^\circ$ skymap
with an integration angle (instrument) of
$\alpha_{\rm int} \!\!= \!\!0.1^\circ$. Actually, a given $l_{\rm crit}$ gives
 $<\!\!n_{\rm cl}\!\!>$ number of clumps: $n_{\rm cl}$ to draw
is obtained from a Poisson distribution of mean value $<\!n_{\rm cl}\!>$.
}{\small
\begin{tabular}{ c  l  l  l r}
\hline
$\!\!\!\!\!$Mass decade$\!\!\!\!\!$ & \# clumps & $l_{\rm crit}$ (kpc) & $<$\# clumps$>$ & $\!\!\!\!\!\!\!\!\!\!$\# to draw$\!\!\!\!\!$\\
  $\log_{10}\left(\frac{M_{\rm cl}}{M_\odot}\right)$ & (Galaxy) & kpc & ($5^\circ\times5^\circ$)& ($5^\circ\times5^\circ$) \\
\hline
   $-6:-5$  &  $3.5 \times 10^{14}$    & $1.0\times 10^{-3}$  &   $1.5\times 10^{-3}$             & 0    \\
   $-5:-4$  &  $4.4 \times 10^{13}$    & $1.0\times 10^{-3}$  &   $1.8\times 10^{-4}$             & 0     \\
   $-4:-3$  &  $5.6 \times 10^{12}$    & $1.0\times 10^{-3}$  &  $2.4\times10^{-5}$               & 0     \\
   $-3:-2$  &  $7.0 \times 10^{11}$    & $7.3\times 10^{-3}$  &  $1.2\times10^{-3}$               & 0    \\
   $-2:-1$  &  $8.8 \times 10^{10}$    & $5.8\times 10^{-2}$  &  $7.6\times10^{-2}$                & 0   \\
   $-1: 0$  &  $1.1 \times 10^{10}$    & $4.0\times 10^{-1}$  &  $2.8$                                     & 3  \\
   $ 0: 1$  &  $1.4 \times 10^{9}$     & $2.0$                &  $3.7\times10^{+1}$                        & 45  \\
   $ 1: 2$  &  $1.7 \times 10^{8}$     & $6.7$                &  $1.0\times10^{+2}$                        & 117  \\
   $ 2: 3$  &  $2.2 \times 10^{7}$     & $1.6\times 10^{1}$   &  $8.7\times10^{+1}$                & 99   \\
   $ 3: 4$  &  $2.8 \times 10^{6}$     & $3.3\times 10^{1}$   &  $3.6\times10^{+1}$                & 44   \\
   $ 4: 5$  &  $3.5 \times 10^{5}$     & $5.8\times 10^{1}$   &  $1.0\times10^{+1}$                & 7    \\
   $ 5: 6$  &  $4.4 \times 10^{4}$     & $9.5\times 10^{1}$   &  $2.2$                                     & 1     \\
   $ 6: 7$  &  $5.6 \times 10^{3}$     & $1.5\times 10^{2}$   &   $4.1\times 10^{-1}$              & 0     \\
   $ 7: 8$  &  $7.0 \times 10^{2}$     & $2.5\times 10^{2}$   &  $6.6\times10^{-2}$                & 0     \\
   $ 8: 9$  &  $8.8 \times 10^{1}$     & $2.9\times 10^{2}$   &  $9.5\times10^{-3}$                & 0     \\
   $ 9:10$  &  $1.1 \times 10^{1}$     & $2.9\times 10^{2}$   &  $1.3\times10^{-3}$               & 0     \\
\hline
\end{tabular}
}
\label{tab:l_crit}
\end{center}
\vspace{-0.5cm}
\end{table}
The flowchart given in Fig.~\ref{fig:flowchart} summarises the various steps
implemented in {\sc clumpy} to obtain the skymap (see, e.g., Fig.~\ref{fig:skymap1}).
\begin{table*}[!t]
\renewcommand{\tabcolsep}{0.47cm}
  \begin{center}
  \caption{Format of the {\tt ASCII} file to describe halos processed by
  {\tt ./bin/clumpy\_dsph -h[option]}. 'Type' should take one of the values of {\tt gENUM\_TYPEHALOES}, namely
{\tt DSPH}, {\tt GALAXY} or {\tt CLUSTER}. 'long','lat' and 'd' are the
  galactic longitude, latitude, and distance to the halo, $z$ its redshift and 
$R_{\rm vir}$ its virial radius. The profile parameters
  are $\rho_s$, $r_s$,  'prof.' (see {\tt profile.h}) and the shape parameters. In particular,
  if 'prof'={\tt kZHAO}, see \S\ref{subsec:parameters}), the profile corresponds to the $(\alpha,\,\beta,\,\gamma)$
  profile (in which case the next-to-last three columns are used). If an {\tt kEINASTO} is used instead, only the first shape parameter
is considered. }
{\small
  \begin{tabular}{ccccccccccccc}
  \hline
  Name& Type & long.&lat. & d  & z& $R_{\rm vir}$   &$\rho_s$             &$r_s$ & prof. & $\alpha$ & $\beta$ & $\gamma$ \\
   -       & -         & deg&deg&kpc&-    & kpc&        $M_\odot$kpc$^{-3}$&kpc&      -   &            - &           - &           -  \\
  \hline
  \hline
  \end{tabular}
}
  \label{tab:dsphs}
  \end{center}
\vspace{-0.5cm}
  \end{table*}

  \paragraph{Threshold mass $m_{\rm thresh}$ for the sub-clumps in an
    ``far-away'' halo}
The second case where the criterion is used is for the sub-clumps
in a halo located away from us, e.g., a dSph galaxy. In that case, the
distance $d$ to the clumps is known 
 as they are all distributed within their parent halo's radius and we
 can determine in a similar fashion the threshold 
mass above which clumps must be drawn.

\section{Description of the code\label{sec:code}}
{\sc Clumpy} is written in C/C++ and relies of the CERN {\tt ROOT} library that is
publicly  available from {\tt http://root.cern.ch}. The main features
of the code are described hereafter along with brief  descriptions of its
most important functions. The code's structure is standard, with separate
directories for the various pieces of code: declarations are
in {\tt include/*.h}, sources in {\tt src/*.cc}, compiled libraries,
objects and executables are respectively in the {\tt lib/}, {\tt obj/},
and {\tt bin/} directories.
\subsection{Code executables \label{subsec:structure}}

The code has one executable that can take three main options
  depending whether the user i) is interested in performing a generic
  analysis on the Galactic halo, ii) is focusing on a list
  of specific halos (e.g. dSph galaxies), or iii) wishes to perform a
  statistical analysis on a single object. These main three usages of the code all
  accommodate several cases, the details of which are given below.

\subsubsection{Galactic mode: {\tt ./bin/clumpy  -g[option]}\label{subsubsec:galactic_mode}} 
This calculates the $J$-factor for the Galactic halo, including both
the smooth and the clump component. For the clump component,
the mean flux or a stochastic realisation can be chosen. Note that a
list of specific halos can be added to the calculation.
Several options allow to select the quantity to be calculated/plotted
from a text interface. A mandatory input 
is the parameter file {\tt clumpy\_params.txt} described in
\S\ref{subsec:parameters} and table~\ref{tab:param}. 
The options are the following: 
\begin{itemize}
\item {\tt 'option = -g1'} plots the total, smooth and averaged clump density in
  1D, given the user-defined total density and the number of clumps in
  a given mass range.
\item {\tt 'option = -g2'} plots $J_{\rm sm} + \langle J_{\rm
    subs}\rangle +  \langle J_{\rm
    cross-prod}\rangle$\footnote{Does not exist for decaying DM.},
  respectively defined by \eq{eq:gal_Jsm}, \eq{eq:gal_meanJcl} and \eq{eq:cross_prod} as a function of the integration angle $\alpha_{\rm int}$.
\item {\tt 'option = -g3'} is the same as {\tt '-g2'} but as a function of the angular distance to the Galactic centre, for a given integration angle.
\item {\tt 'option = -g4'} is the same as {\tt '-g3'} but including a list of
  specific objects, such as dSph galaxies.

\item {\tt 'option = -g5'} plots a 2D skymap of the $J$-factor,  using the averaged clump
  description. where 
\item {\tt 'option = -g6'} is similar {\tt '-g5'}, but adds specific halos to
  the plots (e.g. dSph), to identify their contrast to the background.
\item {\tt 'option = -g7'} and {\tt -g8} are respectively the same as {\tt '-g5'} and {\tt '-g6'} but 
with the clumps randomly drawn from their user-defined mass and spatial distributions. 
\end{itemize}

\subsubsection{Halo list  mode: {\tt ./bin/clumpy -h[option]}} 

This usage of {\sc clumpy} performs operations that are basically the
identical to those performed in the 'Galactic' mode (see
\S\ref{subsubsec:galactic_mode}), but for a list of halos that are
not the Galactic halo. This mode
was initially implemented for the study of dwarf spheroidal galaxies
\citep{2011arXiv1104.0412C}, but can be extended to any DM halo (e.g., galaxy
clusters). If only the smooth total density  profile is considered,
the halo profiles are read from a simple {\tt ASCII} list of profiles
along with their parameters, as exemplified in
Tab.~\ref{tab:dsphs}. If  sub-clumps are considered, the {\sc clumpy}
parameter file {\tt clumpy\_params.txt} must also be provided (see table~\ref{tab:param}). 

Again, further options are accessed for this mode by means of a simple text-user interface: 
\begin{itemize}
\item {\tt '-h1'} to {\tt '-h3'} are identical to {\tt '-g1'} to {\tt
    '-g3'} but transposed to any halo that is not the Galactic DM
  halo;
\item {\tt '-h4'} to {\tt '-h5'} create 2D skymaps of the halo under
  scrutiny using respectively an averaged description or a statistical
  realisation for the sub-clumps within this halo;
\end{itemize}
\begin{table*}
\caption{{\sc clumpy} user-defined input parameter file ({\tt clumpy\_params.txt)}.}
\renewcommand{\tabcolsep}{0.5cm}
\begin{center}
{\small
\begin{tabular}{ll}
\hline
Name & Definition\\
\hline
{\tt gCOSMO\_RHO0\_C} & Critical density of the universe [$M_\odot$~kpc$^{-3}$]\\
{\tt gCOSMO\_OMEGA0\_M} & Present-day pressure-less matter density\\
\vspace{2mm}
{\tt gCOSMO\_OMEGA0\_LAMBDA }& Present-day dark energy density\\
{\tt gDM\_GAMMARAY\_FLAG\_SPECTRUM} & $\gamma$-ray spectrum flag [{\tt gENUM\_GAMMASPECT}]\\

{\tt gDM\_MMIN\_SUBS}  & Min. mass of a DM (sub-)clump [$M_\odot$] \\
{\tt gDM\_MMAXFACTOR\_SUBS}& Defines max. mass of clump in host halo
as $M_{\rm max} = {\rm Factor} \times M_{\rm host}$  \\
\vspace{2mm}
{\tt gDM\_RHOSAT} & Saturation density [$M_\odot$~kpc$^{-3}$]\\
{\tt gTYPE\_CLUMPS\_FLAG\_CVIRMVIR}$^\dagger$&  Clump concentration flag in the parent TYPE halo[{\tt gENUM\_CVIRMVIR}] \\
{\tt gTYPE\_CLUMPS\_FLAG\_PROFILE} & Clump inner profile flag [{\tt gENUM\_PROFILE}]\\
{\tt gTYPE\_CLUMPS\_SHAPE\_PARAMS[0-2]} &Shape parameters for sub-clumps inner profile\\
{\tt gTYPE\_DPDM\_SLOPE}           &  Slope subclump mass function \\
{\tt gTYPE\_DPDV\_FLAG\_PROFILE}   & Spatial distribution subclumps [{\tt gENUM\_PROFILE}]\\
{\tt gTYPE\_DPDV\_RSCALE}      &  Scale radius for the subclump
distribution in the parent halo [kpc] \\
{\tt gTYPE\_DPDV\_SHAPE\_PARAMS[0-2]} & Shape parameters for the clump spatial distribution\\
\vspace{2mm}
{\tt gTYPE\_SUBS\_MASSFRACTION}   &  Mass fraction of the parent halo in sub-clumps \\
{\tt gGAL\_CLUMPS\_FLAG\_CVIRMVIR} & Concentration flag [{\tt gENUM\_CVIRMVIR}] \\
{\tt gGAL\_CLUMPS\_FLAG\_PROFILE} &Clump inner profile flag [{\tt
  gENUM\_PROFILE}]\\
{\tt gGAL\_CLUMPS\_SHAPE\_PARAMS[0-2]} & Shape parameter
for the Galactic clumps inner profile \\
{\tt gGAL\_DPDM\_SLOPE} &Slope the clump mass function\\
{\tt gGAL\_DPDV\_FLAG\_PROFILE} & Clump number distribution flag \\
{\tt gGAL\_DPDV\_RS}              &  Scale radius for clumps [kpc] \\
{\tt gGAL\_DPDV\_SHAPE\_PARAMS[0-2]} & Three shape parameters
for the Galactic clump spatial distribution \\
{\tt gGAL\_SUBS\_M1} & Reference mass for  {\tt gGAL\_SUBS\_N\_INM1M2}
[$M_\odot$] \\
{\tt gGAL\_SUBS\_M2} & Reference mass for {\tt gGAL\_SUBS\_N\_INM1M2} [$M_\odot$] \\
{\tt gGAL\_SUBS\_N\_INM1M2} & \# of clumps in $[M_1,M_2]$ \\
{\tt gGAL\_RHOSOL} & Local DM density [GeV~cm$^{-3}$]\\
{\tt gGAL\_RSOL} & Distance Sun -- Galactic centre [kpc]\\
{\tt gGAL\_RVIR} & Virial radius of the Galaxy [kpc]\\
{\tt gGAL\_TOT\_FLAG\_PROFILE} & Total DM profile of the Milky Way [{\tt gENUM\_PROFILE}]\\
{\tt gGAL\_TOT\_RSCALE}   &  Scale radius for DM halo of the Milky Way [kpc] \\
\vspace{2mm}
{\tt gGAL\_TOT\_SHAPE\_PARAMS[0-2]} & Shape parameter
for the Galactic total density profile\\
{\tt gSIMU\_ALPHAINT\_DEG} & Integration angle [deg]\\
{\tt gSIMU\_EPS}  &  Default precision for numerical integrations\\
{\tt gSIMU\_IS\_ANNIHIL\_OR\_DECAY}  &  For annihilating or decaying DM\\
{\tt gSIMU\_SEED}  &  Seed of random number generator\\
\hline
\end{tabular}
}
\label{tab:param}
{\\\footnotesize$^\dagger$ {\tt TYPE} corresponds to the type of halo under scrutiny (not the Galactic halo): {\tt DSPH}, {\tt GALAXY}
or {\tt CLUSTER} are the values\\ allowed by {\tt gENUM\_TYPEHALOES} in this version.}
\end{center}
\vspace{-0.5cm}
\end{table*}
  \subsubsection{'Statistical' mode:  {\tt ./bin/clumpy -s[option]}}
   Confidence levels on the astrophysical factor $J$, for e.g. dSphs, 
   can also be estimated from an ASCII {\em statistical file}. The format
   for the latter is quite similar to the halo list file (see documentation) as it requires the
   profile parameters, but it differs in the sense that for a single halo, thousands of profiles 'selected' by a statistical analysis
   must be provided (their respective $\chi^2$ value is necessary,
   although not used for all CL calculations). Such a statistical
   analysis can be the Markov Chain Monte Carlo (MCMC), as discussed
   in Charbonnier et al. \citep{2011arXiv1104.0412C}; Walker et al. \citep{2011ApJ...733L..46W}. The most important options in the statistical mode are:
  \begin{itemize}
     \item {\tt '-s2'} plots the probability density functions (PDF) and
      correlations of the parameters given in the statistical
      file;
    \item {\tt '-s4'} locates the $x\%$ confidence intervals on the PDF
     of each parameters;
    \item {\tt '-s5'} computes, for each radius from the dSph centre, the
      median and confidence levels of the DM density $\rho(r)$;
    \item {\tt '-s6'} is the same as {\tt '-s5'} but for $J(\alpha_{\rm
int})$, i.e. as a function of the integration angle;
    \item {\tt '-s7'} is same as {\tt '-s5'} but for $J(\theta)$,
      i.e. as a function of the angle to the centre of the halo. 
    \end{itemize}

\subsection{User-defined parameters \label{subsec:parameters}}

For most of the options used above, a user-define parameter file is read.
Table~\ref{tab:param} gives the list of the main parameters. 
Each parameter is a single word formatted as {\tt g} (for global parameter) + physics
keyword + parameter name. For instance, {\tt COSMO} refers to
cosmology-related parameters, {\tt TYPE} defines the type of halo under scrutiny
(dSPh galaxy, galaxy or galaxy clusters) and {\tt GAL} refers to the Milky Way halo parameters. Note also that the parameters ending in {\tt FLAG\_CVIRMVIR}, {\tt FLAG\_SPECTRUM}, and {\tt FLAG\_PROFILE} should take values from the following enumerations:
\begin{itemize}
\item {\tt  enum  gENUM\_CVIRMVIR   \{kB01, kENS01, kJS00\}} for the clump concentration;
\item {\tt  enum  gENUM\_GAMMASPECT \{kSUSY\_BUB98, kSUSY\_TO02, kSUSY\_BBE08\}} for the $\gamma$-ray spectrum;
\item {\tt  enum  gENUM\_PROFILE    \{kZHAO, kEINASTO, kEINASTO\_N\}}
  for the DM spatial distributions (smooth and clumpy components). The
  corresponding shape parameters (e.g. ($\alpha, \beta, \gamma$) for
  {\tt kZHAO}, $\alpha$ for {\tt kEINASTO} and $n$ for {\tt
    kEINASTO\_N}) should be provided by the user in the input parameter
  file (see below). 
\end{itemize}
This enumerations allow the user to use their explicit denomination {\tt kXXX},
rather than the corresponding integer number.
These parameters are gathered in the ASCII file {\tt clumpy\_params.txt} which  is formatted as
``Parameter name'', ``Units'', ``Value''. The function {\tt
load\_parameters(file\_name)} defined in {\tt params.h} reads in the input file and
assigns the appropriate values to the parameters. The user can change
any ``Value" in this file\footnote{Some of these parameters are optional
and used only for specific {\tt modes[options]}}.

\subsection{{\sc clumpy} libraries\label{subsec:libraries}}

The functions of {\tt {\sc clumpy}} are located in eleven ``library'' files, accordingly
to their field of action. We briefly describe hereafter the most important 
functions of these libraries. We are omitting in this document the more minor
functions, the description of which can be found in the documentation 
provided with the code. Also, we do not detail here what arguments are passed 
in these functions as it can be easily retrieved from the documentation.

\subsubsection{\tt integr.h \label{subsubsec:libintegr}} 
This library contains standard integration algorithms. The Simpson integrator with
doubling step is the most widely used throughout the code to ensure that the 
user-defined precision is reached. Depending on the 
function to integrate, either linear or logarithmic stepping can be used.

\subsubsection{\tt geometry.h \label{subsubsec:libgeom}}
In {\tt geometry.h} are gathered functions performing geometrical tasks (change of coordinates, etc.). 
They are not essential for the presentation of the code and generally self-explanatory. 
We refer the reader to the extensive Doxygen documentation and to \ref{app:geom} for more details.

\subsubsection{{\tt inlines.h} and {\tt misc.h}} These contain some definitions (unit conversion and inline
functions) and miscellaneous formatting functions respectively.

\subsubsection{\tt profiles.h \label{subsubsec:libprofiles}}
This library contains all the density profiles available in the code
and operations on those. Among the functions in the ``profiles''
section of the code, some act in a very straightforward way
(e.g. squaring the density) and are not included in the present
document. The reader is, once again, referred to the Doxygen
documentation for a full description.

\begin{itemize}
  \item {\tt rho\_ZHAO} returns the DM density of a user-defined profile having
  the generic functional form given in Zhao \citep{1996MNRAS.278..488Z}
  \beq
  \rho(r) ^{\rm ZHAO}=\frac{\rho_s}{\left(r/r_s\right)^\gamma\,\left[1+(r/r_s)^\alpha\right]^{(\beta-\gamma)/\alpha}}\;.
  \label{eq:generic}
  \eeq
The isothermal, Navarro et al. \citep{1997ApJ...490..493N},
Moore et al. \citep{1998ApJ...499L...5M},  and Diemand et al. \citep{2004MNRAS.353..624D} profiles are all from the Zhao family.
They are defined using \eq{eq:generic} with the set of parameters of
table~\ref{table:profile_param}. The user is required to specify these
parameters in the ASCII input file.

\begin{table}[!t]
\begin{center}
\caption{Standard DM density profiles parameters, following \eq{eq:generic}.}
\renewcommand{\tabcolsep}{0.4cm}
{\small
\begin{tabular}{l c c c l}
\hline
  & $\alpha$ & $\beta$ & $\gamma$ & Reference \\
\hline
ISO & 2 & 2 & 0 & - \\
NFW97 & 1 & 3 & 1  & Navarro et al. \citep{1997ApJ...490..493N}\\
M98 & 1.5 & 3 & 1.5  & Moore et al. \citep{1998ApJ...499L...5M}\\
DMS04 & 1 & 3 & 1.2  & Diemand et al. \citep{2004MNRAS.353..624D}\\ 
\hline
\end{tabular}
}
\label{table:profile_param}
\vspace{-0.5cm}
\end{center}

\end{table}
\item {\tt rho\_EINASTO} and {\tt rho\_EINASTO\_N}: it has been argued that a profile with a logarithmic inner slope
  gives a better fit to the simulated data. This so-called Einasto is
  generically written as
\[
\rho(r)\propto\exp\left(-Ar^\alpha\right)\;.
\] 
In Navarro et al. \citep{2004MNRAS.349.1039N} and Springel et al. \citep{2008MNRAS.391.1685S} the Einasto profile is defined as

\beq
\rho^{\rm EINASTO}(r)= \rho_{-2}\exp\left(-\frac{2}{\alpha}
      \left[\left(\frac{r}{r_{-2}}\right)^{\alpha}-1\right]\right)\;,  
\eeq
where $r_{-2}$ is the radius at which the profile's slope
$d\ln\rho/d\ln r = -2$, and $\rho_{-2}$ the corresponding
density. Both Navarro et al. \citep{2004MNRAS.349.1039N} and Springel
et al. \citep{2008MNRAS.391.1685S} find $\alpha\sim 0.17$ to be a good
fit to the smooth component of simulated halos. Springel et al. \citep{2008MNRAS.391.1685S} further notes that $\alpha \sim 0.68$ allows a good description of the spatial distribution of sub-structures $dP/dV$.

Merritt et al. \citep{2006AJ....132.2685M} use an Einasto $r^{1/n}$ profile, slightly different from the above and  given by
\beq
\rho^{\rm EINASTO\_N}(r)=\rho_e \exp\left(-d_n\cdot \left[\left(\frac{r}{r_e}\right)^{1/n}-1\right]\right)\;.
\eeq
where $n=6$ and $d_n=3n-1/3+0.0079/n$. The scale radius $r_e$ in this relation correspond to the radius within which half of the halo mass is enclosed.

\item The normalisation density of all these profiles ($\rho_s$,
  $\rho_{-2}$ or $\rho_e$) can be done by i) either requesting a
  given value for the density at a given point or ii) by requesting a
  given mass within a given radius. The functions {\tt
    set\_par0\_given\_rhoref} and {\tt
    set\_par0\_given\_mref} achieve the normalisation using both
  approaches.

\item {\tt rho} simply returns the value of the density by calling any
  of the functions above, depending on the value of the profile flag
  given in {\tt clumpy\_params.txt}.

\item For two density profiles $\rho_1$ and $\rho_2$, {\tt rho\_mix}
  returns either the difference $\rho_1 - \rho_2$ or the product
  $\rho_1\rho_2$. This is needed when computing \eq{eq:rhosm} for the smooth density
  profile or \eq{eq:cross_prod} for the cross product.  

\end{itemize}

\subsubsection{\tt integr\_los.h\label{subsubsec:libintegrlos}}
This is the core of the {\sc clumpy} code as it contains the routines performing the integration 
along the line of sight direction ($\theta, \psi$) and over the chosen angular resolution $\alpha_{\rm int}$. 
Given a function $g(l,\alpha,\beta; \theta, \psi)$ where ($l, \alpha, \beta)$ are spherical coordinates
in the $(\theta, \psi)$ direction (see Fig.~\ref{fig:geom}), the integration along
the observation cone, from $l_1$ to $l_2$ is generically written as:
\beq
I(\theta, \psi)=\int_0^{2\pi}d\beta \int_0^{\alpha_{\rm int}} \sin\alpha\, d\alpha 
\int_{l_1}^{l_2} g(l,\alpha,\beta;\theta,\psi)\, l^2 dl\;.
\label{eq:I_theta_psi}
\eeq

\begin{itemize}
\item {\tt integrand\_l}: returns the value of the function 
$f(l,\alpha,\beta; \theta, \psi) \equiv g(l,\alpha,\beta; \theta, \psi)\, l^2$.
A switch is included to integrate different functions of the density
profile $\rho$. Note that here $\rho$ can refer to any profile or
distribution, be it the density or the spatial distribution of clumps
$dP_V/dV$ depending on the normalisation used. The functions that can
be integrated along the line of sight are:
\begin{itemize}
\item $f(l,\alpha,\beta; \theta, \psi) = \rho$ when calculating the
  clumps averaged contribution $\langle J_{\rm subs}\rangle$ of \eq{eq:gal_meanJcl}.
\item $f(l,\alpha,\beta; \theta, \psi) = \rho^2$ in order to evaluate
  J such as in \eq{eq:gal_Jsm}.
\item $f(l,\alpha,\beta; \theta, \psi)=l^2\rho$ is used to calculate
  the mass fraction or number fraction of clumps in the integration volume;
\item $f(l,\alpha,\beta; \theta, \psi)=l^3\rho$ is needed when
  estimating the mean distance of the clumps as in \eq{eq:mean_dist}.
\item $f(l,\alpha,\beta; \theta, \psi)=l^4\rho$ is needed when
  estimating the variance of distance of the clumps.
\item $f(l,\alpha,\beta; \theta, \psi)=l^{-2}\rho$ must be evaluated
  to evaluate the variance on J as in \eq{eq:var_J}.
\end{itemize}
Note that if decaying rather than annihilating DM is being considered, i.e. {\tt gDM\_IS\_ANNIHIL\_OR\_DECAY}
is {\tt false} instead of {\tt true}, any  $\rho^2$ is
replaced by $\rho$ in the above functions\footnote{This makes some of these functions
redundant for decay, but they are still required for annihilation.}.
 
A variation of this function, called {\tt integrand\_lrel}, is used whenever the steepness of the profile
requires it (see Doxygen documentation for more details).

\item {\tt fn\_beta\_alpha}: computes the third integrand in \eq{eq:I_theta_psi}, namely the integration over $l$.
The functions to integrate roughly behave as power laws, with a varying slope, and log-step Simpson 
integration schemes are used (see {\tt integr.h}). Some profiles can be very steep and several tests and tricks are 
needed to ensure the numerical stability/optimisation of the integration. 
The reader is again referred to the documentation for an explicit description of the integration strategy.

\item {\tt fn\_beta}: computes the integration over the angle $\alpha$ in \eq{eq:I_theta_psi}. 
A simple linear Simpson integration scheme is used.

\item {\tt los\_integral}: it is the final part of the triple integration, i.e. the integration over 
the angle $\beta$. This is again performed quite easily with a linear
Simpson scheme. This function is called whenever a single density
profile is to be dealt with.

\item {\tt los\_integral\_mix}: as seen previously, it may be
  necessary to integrate a difference or a product of two
  densities along the line of sight. A flag parameter {\tt
    switch\_rho} is introduced to tell {\tt rho\_mix} to return either
  $\rho_1-\rho_2$ ({\tt switch\_rho = 0})  or $\rho_1\rho_2$ ({\tt switch\_rho = }1) (see {\tt profiles.h}). This
  function then behaves identically to {\tt los\_integral} and uses
  the same flags with respect to the quantity to integrate.

\end{itemize}

\subsubsection{\tt clumps.h \label{subsubsec:libclump}}
All the functions related to the clumps mass and spatial distributions are gathered in this
file. Here, we only give an overview of the most important ones.
\begin{itemize}

\item{\tt add\_halo\_in\_map}: when using the code to generate a
  skymap where clumps are randomly drawn from their mass and spatial distribution,
  this adds the J contribution of the drawn clump to the appropriate
  pixels. The clump is only added to a pixel if $J_{\rm cl}({\rm
    pixel}) > \epsilon \times J_{\rm sm}({\rm pix}_0)$ where ${\rm pix}_0$
  is the pixel containing the centre of the clump and $\epsilon$ is the requested integration precision {\tt gSIMU\_EPS}. 
This avoids the clump from being integrated ``too far'' which would only be time
  consuming without changing the result.

\item {\tt dpdv\_XXX} and {\tt set\_normprob\_dpdv}: the former function
  returns the spatial distribution of the clumps as described in
  \S\ref{subsubsec:clumps} and the second calculates the appropriate
  normalisation to make it a probability according to
  \eq{eq:normalisation-clumps}. The function {\tt dpdv\_XXX} is written in
  several versions depending of the reference frame and coordinate system
  in which the clumps are drawn (cartesian {\tt XXX=xyz}, spherical
  centred on the galactic centre {\tt XXX=rthetapsi}, spherical centred
  on Earth {\tt XXX=lthetapsi}, or centred on Earth, in the cone
  $\Delta\Omega$, in the direction $(\psi, \theta)$ {\tt XXX=lalphabeta}). 

\item {\tt dpdm} and {\tt set\_normprob\_dpdm}: this is the equivalent of the above
but for the mass distribution of the clumps as given by \eq{eq:clump_mass_function}.

\item {\tt frac\_nsubs\_in\_foi} returns the number fraction of clumps
    in a given field of integration w.r.t. the total
    number of clumps in the host halo

\item {\tt frac\_nsubs\_in\_m1m2} returns the fraction of clumps with
  a mass in the range $[m_1,m_2]$, given the total number of clumps.

\item {\tt jcrossprod\_continuum} returns the value of the cross-product integrated along the l.o.s
as defined by \eq{eq:cross_prod}. This is
  a simple call to {\tt los\_integral\_mix}. 

\item {\tt jcrossprod\_interpolation} interpolates the value of $J$ in the pixels
of a skymap, provided that $J$ has previously been estimated in some locations of
the map. This makes use of ROOT interpolation function based on Delaunay's
triangles  and allows to speed-up the calculation significantly when the skymap
has a (too) large number of pixels.

\item {\tt j\_smooth} returns the $J$-factor when a single density profile is used
  (e.g. $\rho_{\rm tot}$). This is a simple call to {\tt los\_integral}.

\item {\tt j\_smooth\_mix} is the same as above but deals with a
  combination of densities, e.g. when computing \eq{eq:gal_Jsm} where
  $\rho_{\rm sm}=\rho_{\rm tot}-\langle \rho_{\rm cl}\rangle$. This is
  a simple call to {\tt los\_integral\_mix}.

\item {\tt j\_sub\_continuum} computes and returns the averaged
  $J$ generated by the sub-clumps in a  halo, in a given line of
  sight.

\item {\tt lum\_singlehalo} returns the total intrinsic luminosity of
  a DM halo, given its inner density profile as
  in \eq{eq:luminosity}.

\item {\tt mass\_singlehalo} returns the total mass of
  a DM halo, given its inner density profile.

\end{itemize}

\subsubsection{\tt spectra.h \label{subsubsec:libpp}}
This library contains the different parametrisations of the average $\gamma$-ray spectra, 
\beq
\frac{\ud N_\gamma}{\ud E_\gamma} \equiv \sum_f \frac{\ud N^f}{\ud E_\gamma} B_f
\eeq
in the particle physics term of \eq{eq:term-pp}.
As mentioned in \S\ref{subsec:pp-term}, we have included the
Bergstr\"om et al.\citep{1998APh.....9..137B} and Tasitsiomi \& Olinto
\citep{2002PhRvD..66h3006T} parametrisations for the $\gamma$-ray continuum and the 
Bringmann et al. \citep{2008JHEP...01..049B} expression of the internal bremsstrahlung. 
Defining $x\equiv E_\gamma/m_\chi$, the three parametrisations are given by their corresponding 
functions below.
\begin{itemize}
\item {\tt dNdE\_BERGSTROM} returns 
\[
\frac{\ud N_\gamma}{\ud E_\gamma}=0.73\cdot\frac{1}{m_\chi}\cdot\frac{\exp\;(-7.8x)}{x^{1.5}}\;.
\]
\item {\tt dNdE\_TASITSIOMI} computes the $\gamma$-ray spectrum as
\[
\frac{\ud N_\gamma}{\ud E_\gamma}=\frac{1}{m_\chi}\cdot\left(\frac{10}{3}-\frac{5}{4}x^{0.5}-\frac{5}{2}x^{-0.5}
+\frac{5}{12} x^{-1.5}\right)\;.
\]
\item {\tt dNdE\_BRINGMANN} estimates the internal bremsstrahlung contribution with
\[
\frac{\ud N_\gamma}{\ud E_\gamma}=\frac{1}{m_\chi}\cdot \frac{\alpha}{\pi}\cdot
\frac{4\left(x^2-x+1\right)^2}{x\left(1-x+\epsilon/2\right)}\cdot
\left[\ln\left(2\cdot\frac{1-x+\epsilon/2}{\epsilon}\right)-x^3+x-\frac {1}{2}\right]\;.
\]
where $\alpha=1/137$ is the fine structure constant and $\epsilon$ is the ratio of the mass of the 
$W$~boson to the neutralino mass $\epsilon=m_W/m_\chi$.
\end{itemize}
\begin{figure*}[!t]
  \begin{center}
\includegraphics[clip=, width=0.48\textwidth]{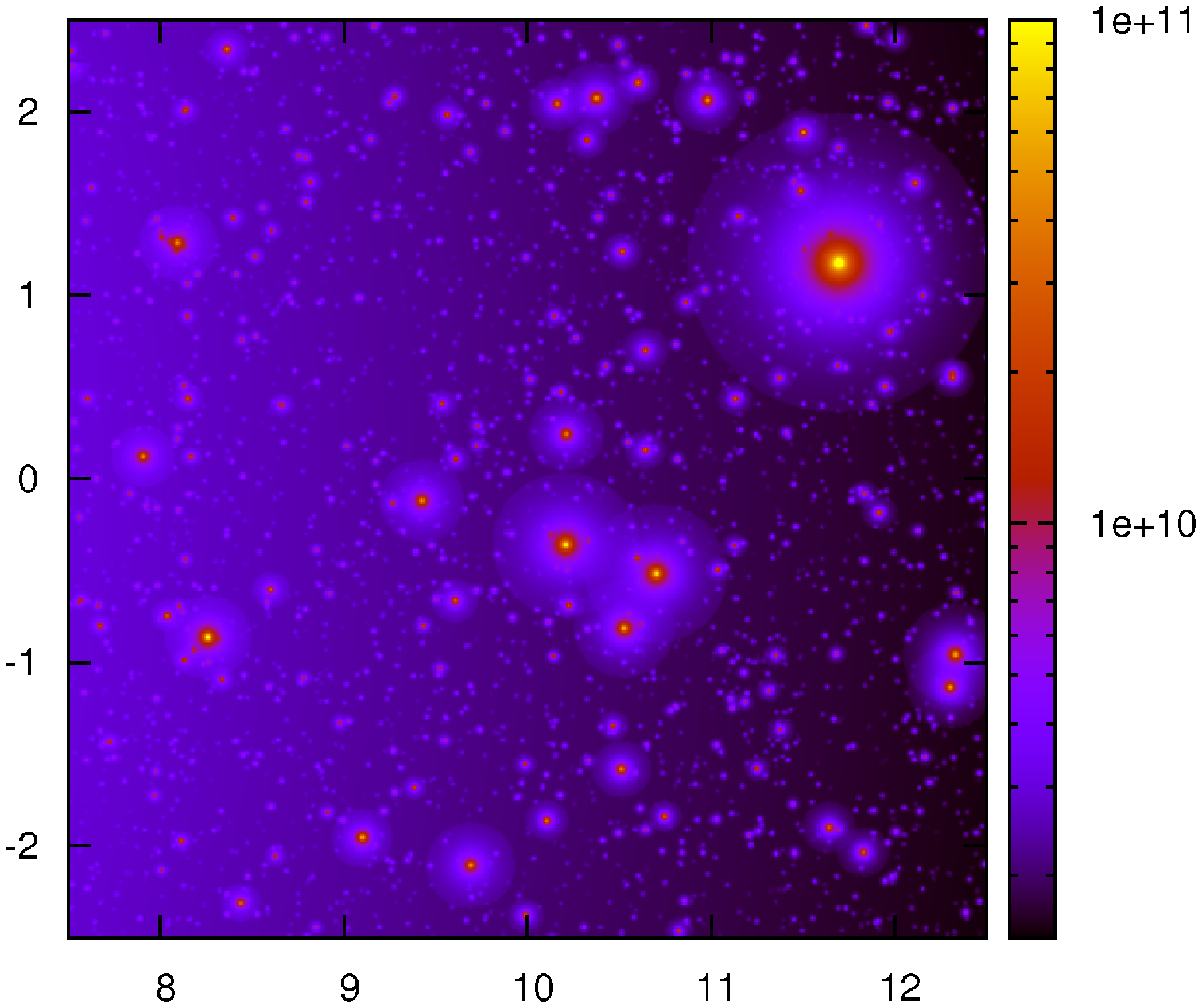}
\includegraphics[clip=, width=0.48\textwidth]{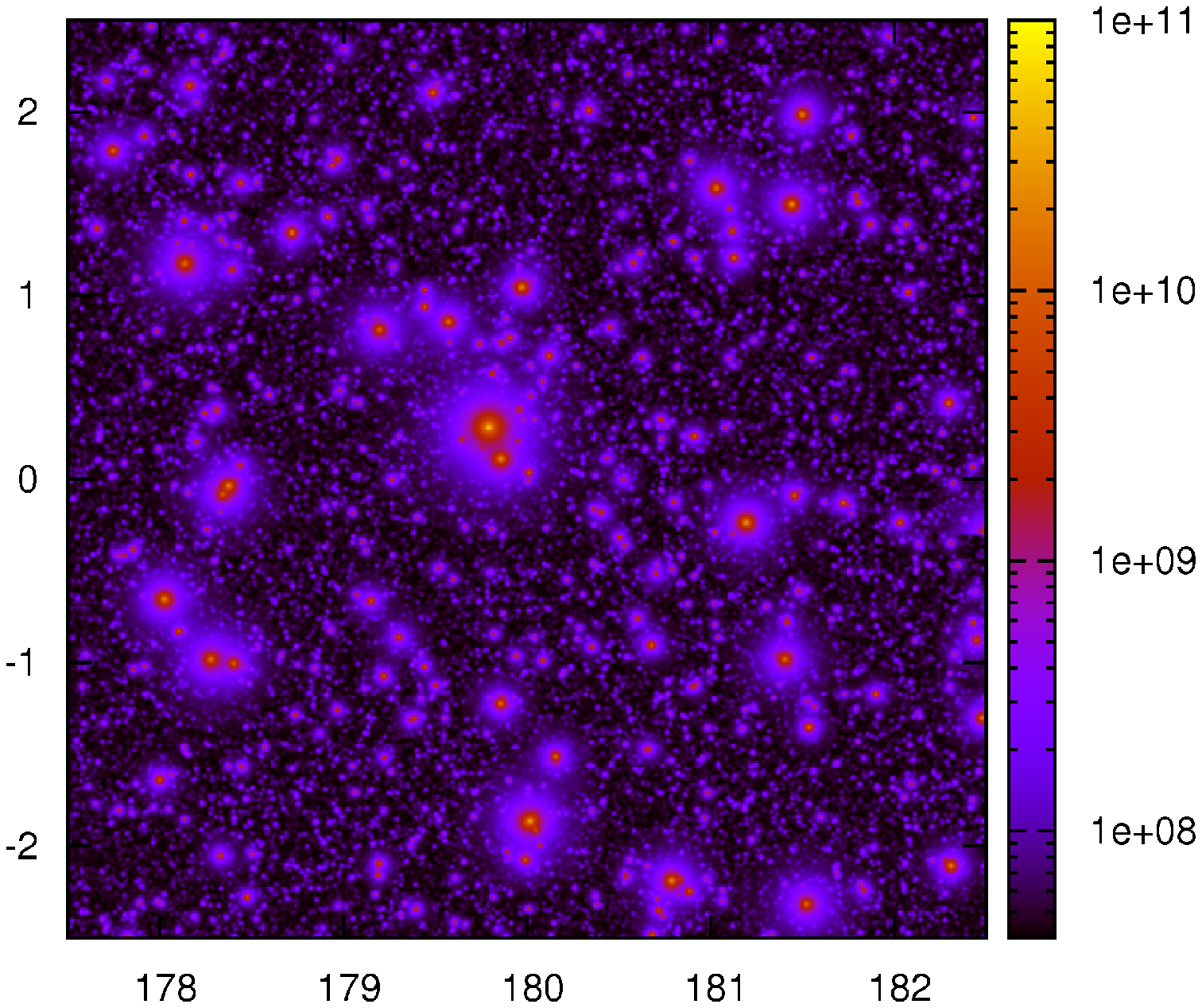}
\caption{Example of two $5^{\circ}\times 5$ skymaps, at $10^\circ$ from the Galactic centre 
(left) and towards the anti-centre(right), for a $0.01^\circ$ angular resolution, 
obtained from the skymap mode of {\sc clumpy}. The colour scale gives the $J$ values.
These maps are obtained with the following command option:
{\tt ./bin/clumpy  -g7}. {\sc Clumpy} will display the results automatically but these images have been produced with 
Gnuplot for aesthetic reasons. 
\label{fig:skymap1}}
\end{center}
\vspace{-.5cm}
\end{figure*}
\subsubsection{\tt stat.h \label{subsubsec:libstat}}
The functions included in this library perform simple statistical
operations on the random variables of the problem (mass and distance
of the clumps) and on quantities deriving from them (luminosity,
J-factor). Ultimately, these are mainly used to determine the critical
distance (or threshold mass) below (above) which clumps need to be
drawn (see
\S\ref{subsubsec:l_crit} for the adopted procedure). Given these dark matter
spatial ($d{\cal P}_V/dV$) and mass ($d{\cal P}_M/dM$) distributions defined in \S\ref{subsubsec:clumps}):
\begin{itemize}
\item  for a given mass and/or distance range, {\tt mean1cl\_mass},  {\tt mean1cl},  {\tt
    mean1cl\_lum2}, {\tt mean1cl\_l},  and {\tt
    mean1cl\_j} return respectively the average mass, luminosity, luminosity squared,
  distance and J-factor of one clump;
\item {\tt var1cl\_mass},  {\tt var1cl\_lum},  {\tt var1cl\_l},  and {\tt
    var1cl\_j} returns the variance of the corresponding random
  variable;
\item finally, knowing the variance of the J-factor of one clump, {\tt
  find\_lcrit\_los} and {\tt find\_mthresh\_los} find the critical distance and
threshold mass for drawing the clumps. They both rely on a simple
dichotomy to bracket the desired quantity, as illustrated on Fig.\ref{fig:RE}.
\end{itemize}
\subsubsection{\tt janalysis.h \label{subsubsec:libdsph}}
This is the top-level library. It contains the main functions called when
invoking any of the three options of clumpy (galactic halo, halo list or
statistics, see \S\ref{subsec:structure}). Note that the first word
'{\tt xxx}'
of the function's name refers to which option they belong to ({\tt xxx=gal},
{\tt xxx=halo}, or {\tt xxx=stat}). 

\begin{itemize}
\item  {\tt xxx\_j1D} handles all the 1D calculations of the
  Galactic or halo options. It computes $\rho(r)$,
   $J(\alpha_{\rm int})$, and $J(\theta)$, where $\theta$ is the angle measured
   w.r.t. the Galactic centre.
\item  {\tt xxx\_j2D} generates the 2D skymaps for the smooth,
  averaged and statistically drawn clump contributions. It also
  include the cross product term and It applies to
   both the Galactic and halo options.
\item {\tt xxx\_load\_list} loads the list of halo under scrutiny
  either for the halo mode or the statistical option. The list should
  be provided as an ASCII file with the format given in table~\ref{tab:dsphs}.
\item  {\tt xxx\_set\_par*} are a set of functions filling the
  parameters arrays to be passed to the above functions with the
  proper quantities (e.g., profile flag and shape parameters,
  concentration flag, etc.). It applies to the Galactic  and halo options.
\item {\tt stat\_CLs} is the most important function in the
  statistics option and computes the confidence levels of a quantity
  (density, J-factor) from its PDF.
\end{itemize}
\begin{figure*}[!t]
  \begin{center}
\includegraphics[clip=,width=0.32\textwidth]{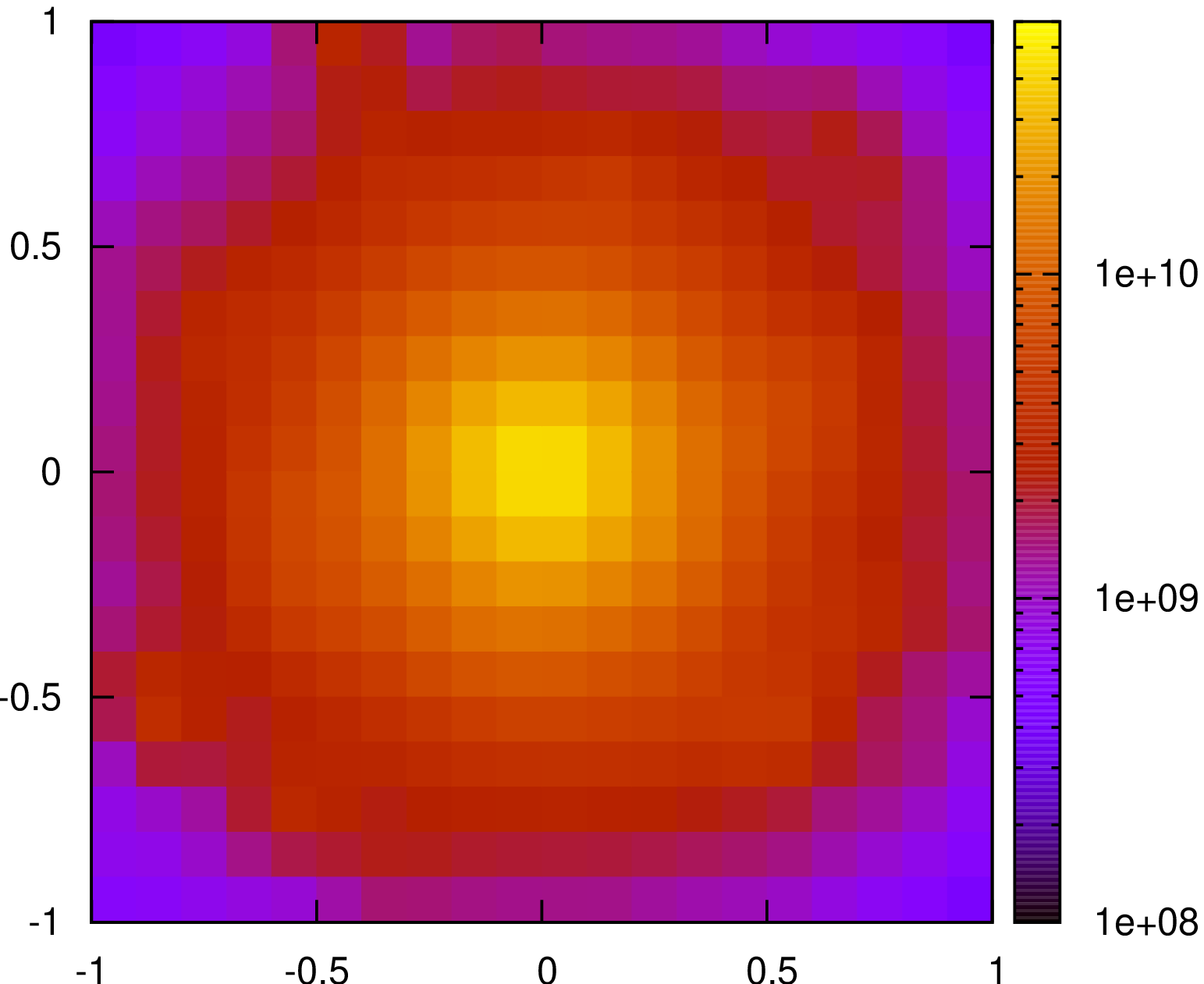}
\includegraphics[clip=, width=0.32\textwidth]{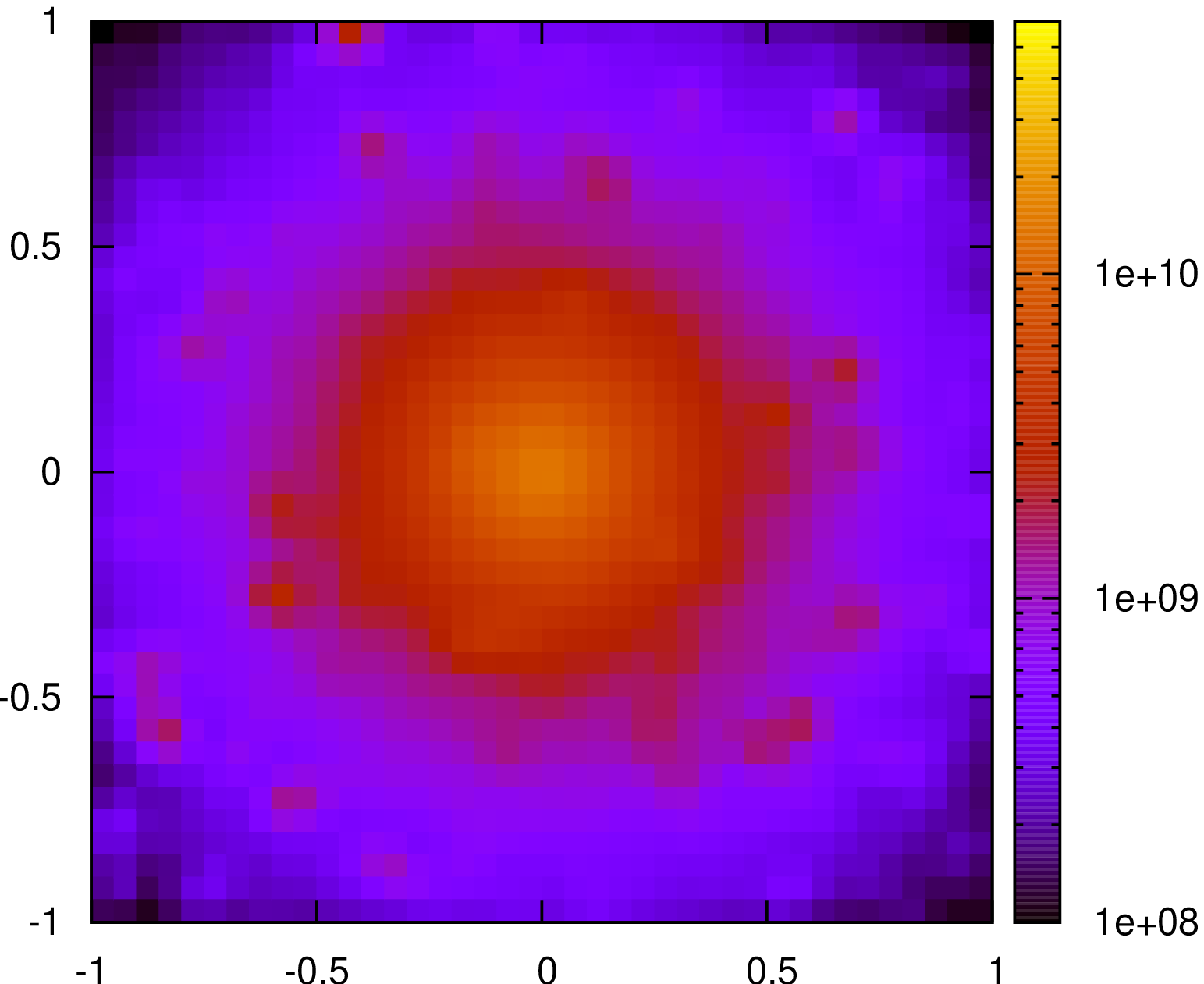}
\includegraphics[clip=, width=0.32\textwidth]{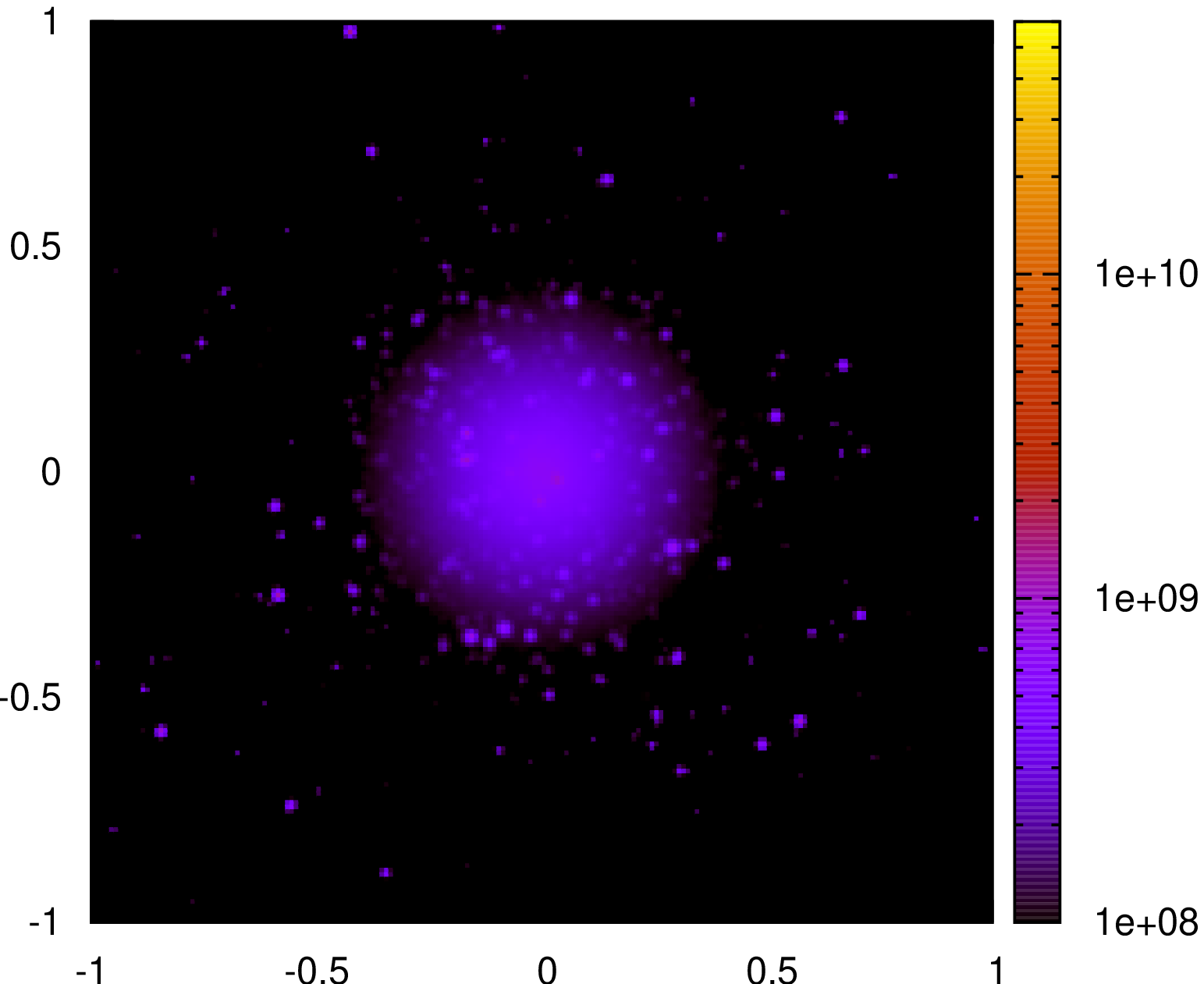}
\caption{Skymaps of a fake dSph galaxy at three different
  resolutions: $0.1^\circ$ (left), $0.05^\circ$ (middle), and
  $0.01^\circ$(right). Note that ROOT will automatically display the image, but these ones
  have been plotted using the output files with Gnuplot.
These maps are obtained with the following command option:
{\tt ./bin/clumpy -h5}
\label{fig:skymap2}}
\end{center}
\vspace{-0.5cm}
\end{figure*}

\section{Run examples \label{sec:examples}}

This section provides a few plots obtained by using the
{\tt clumpy} executable.
Many more examples and illustrations are provided along with
the Doxygen documentation.

\subsection{Sky maps\label{subsec:sky_maps}}

Figure~\ref{fig:skymap1} shows two examples of a $5^\circ\times 5^\circ$
skymap of the J-factor (option {\tt '-g7'} of {\tt clumpy\_gal}), when looking $10^\circ$ off the Galactic
centre (left) or towards the anti-centre (right). The angular
resolution is $0.01^\circ$ and the Galactic smooth DM density, 
clump distribution and clump inner profile are all assumed to follow
an NFW parametrisation.

Similarly, skymaps of individual objects can be performed (option {\tt '-h5'} of
{\tt clumpy\_dsph}). In Fig.~\ref{fig:skymap2}, a "generic'' dwarf galaxy,
including a population of subclumps, is integrated over $0.1^\circ$ (left),
$0.05^\circ$ (middle), and $0.01^\circ$ (right). As the resolution increases, more
substructures become apparent but the J-factor is strongly reduced.

\subsection{dSphs and statistical tools\label{subsec:{dwarf}}}

\begin{figure*}[!th]
 \begin{center}
\includegraphics[clip=, width=0.45\textwidth]{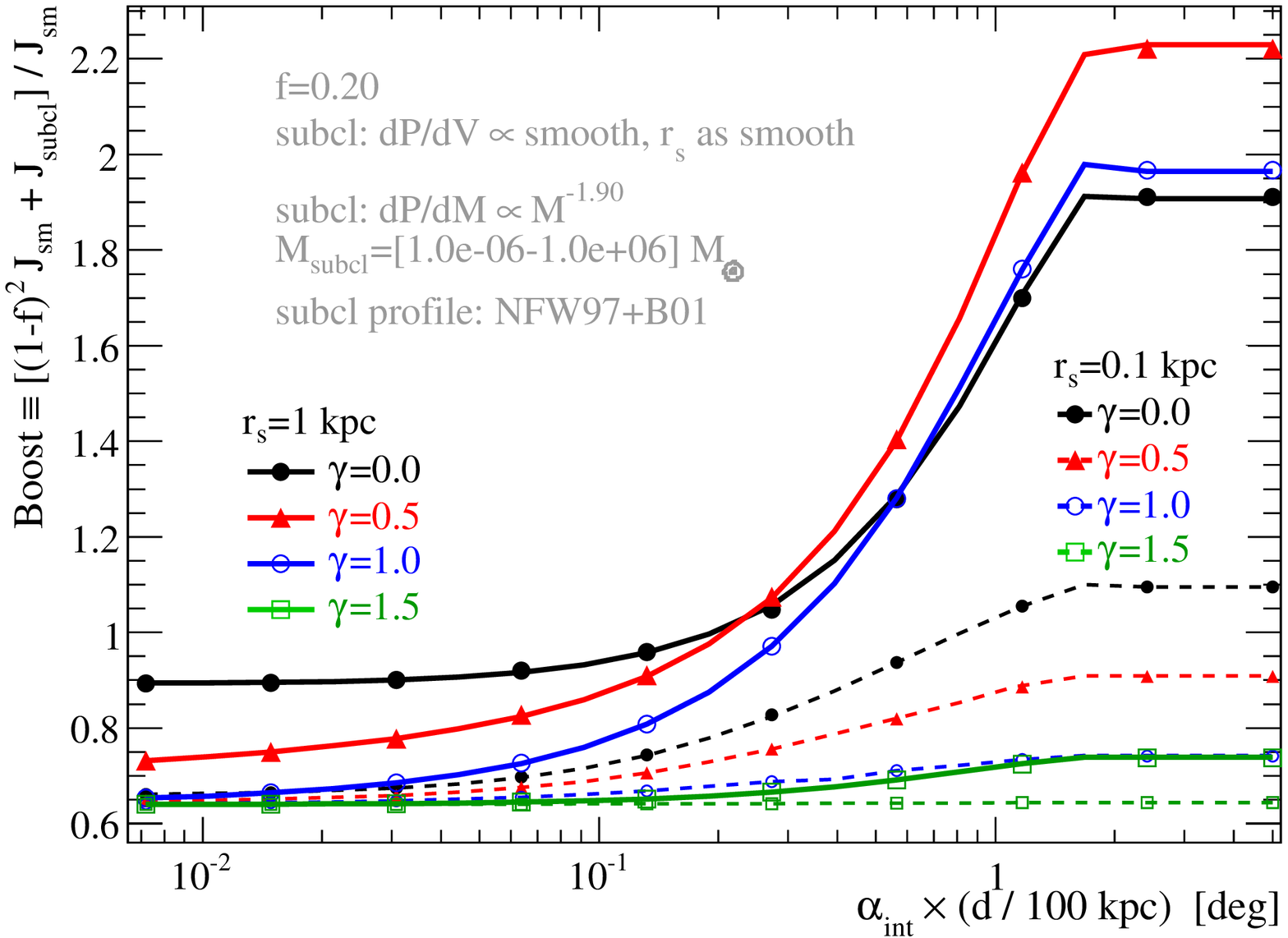}
\includegraphics[clip=, width=0.45\textwidth]{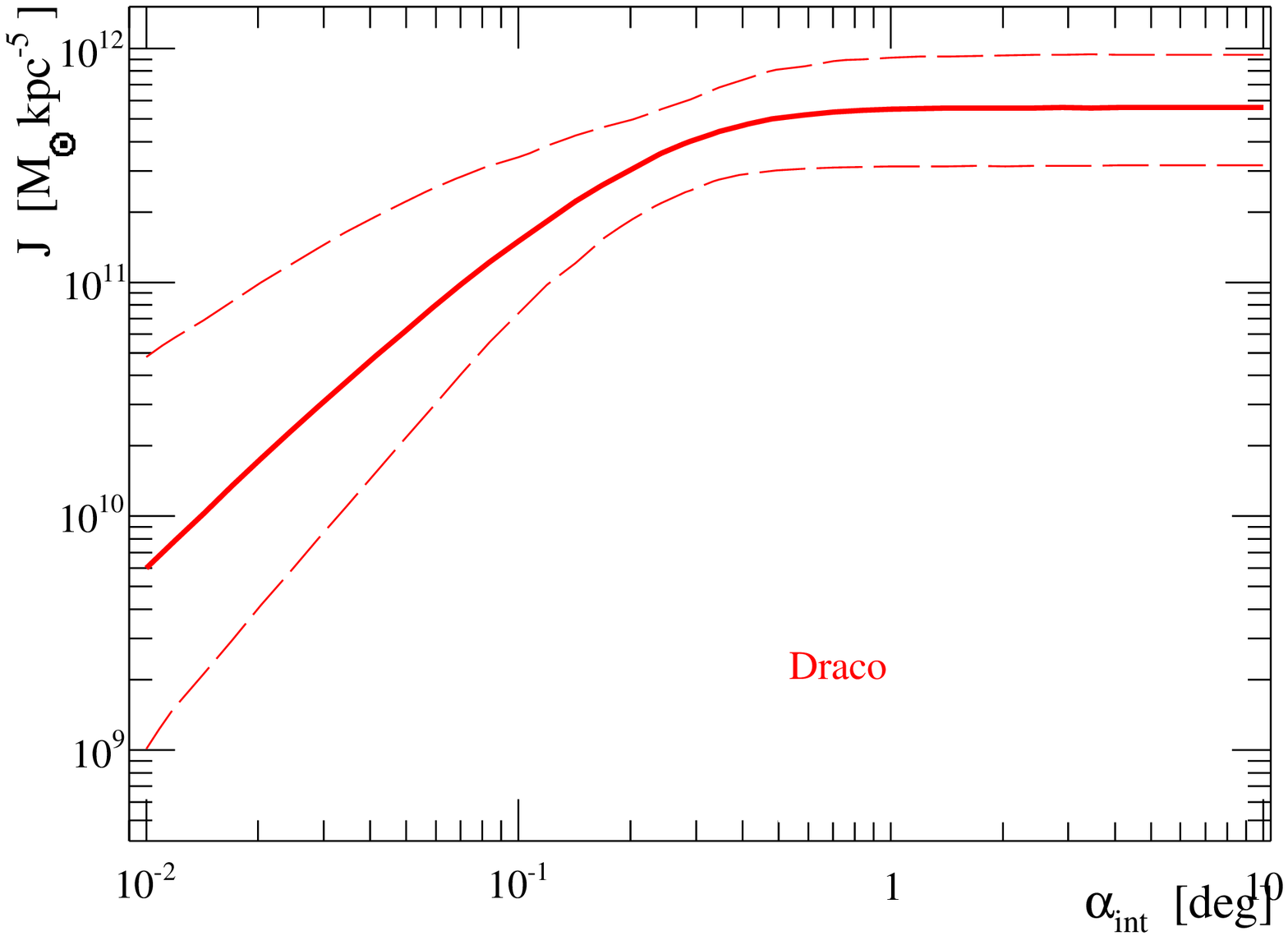}
\includegraphics[clip=, width=0.45\textwidth]{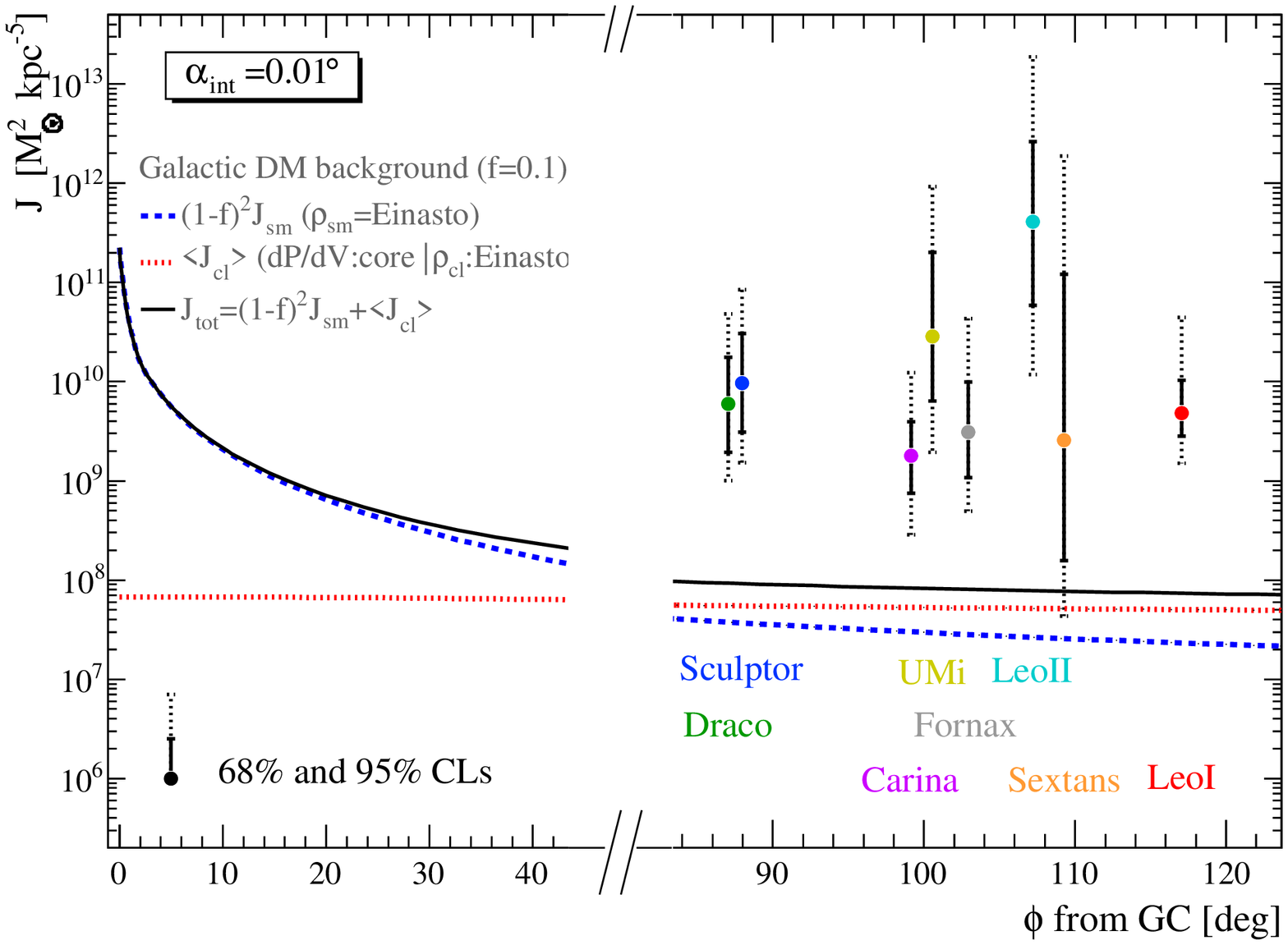}
\caption{{\bf Top left:} boost factor as a function of the integration angle
for 'several' generic dSphs ({\tt ./bin/clumpy -h2}). {\bf Top right:} median (solid line) and
95\% confidence levels (dashed lines) of the $J$-factor as a function
of the integration angle for the Draco dSph ({\tt ./bin/clumpy -s6}). {\bf Bottom panel:}
the $J$-factor for the dSph can be over-plotted on the galactic
smooth and mean-clump $J$ factors ({\tt ./bin/clumpy -g4}).\label{fig:CLs}}
\end{center}
\vspace{-0.5cm}
\end{figure*}

The three panels in Fig~\ref{fig:CLs} illustrate {\sc clumpy}
capabilities on the dSphs. The top left and bottom figures are
taken from  Charbonnier et al. \citep{2011arXiv1104.0412C}, and the top right one is a direct
result of the analysis performed in the above paper. Running
{\sc clumpy} on the default parameters will not give these
plots, but similar ones in spirit.

\begin{figure*}[!t]
\begin{center}
\includegraphics[width=0.65\textwidth]{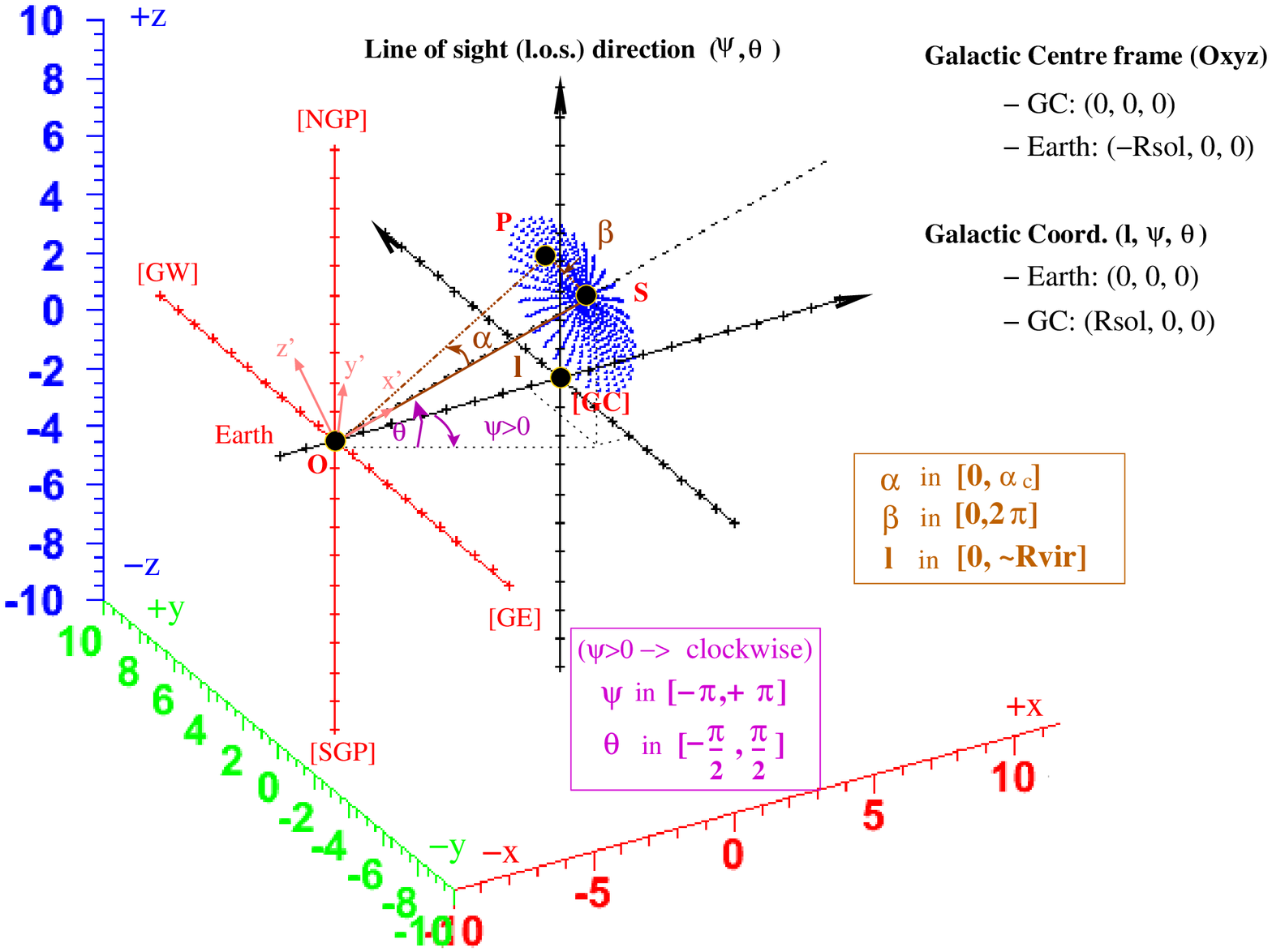}
\end{center}
\caption{The geometry as defined in the {\sc clumpy} simulation. Two frames are
used: the Galactic Centre frame and the Galactic coordinates.  For a
given direction of observation $\left(\psi, \theta \right)$ (l.o.s. for
line of sight), we will run over all the $\alpha$ and $\beta$  directions
within the solid angle $\Delta\Omega$.}
\label{fig:geom}
\vspace{-0.5cm}
\end{figure*}

\begin{figure*}[!t]
\centering
\includegraphics[width=0.31\textwidth]{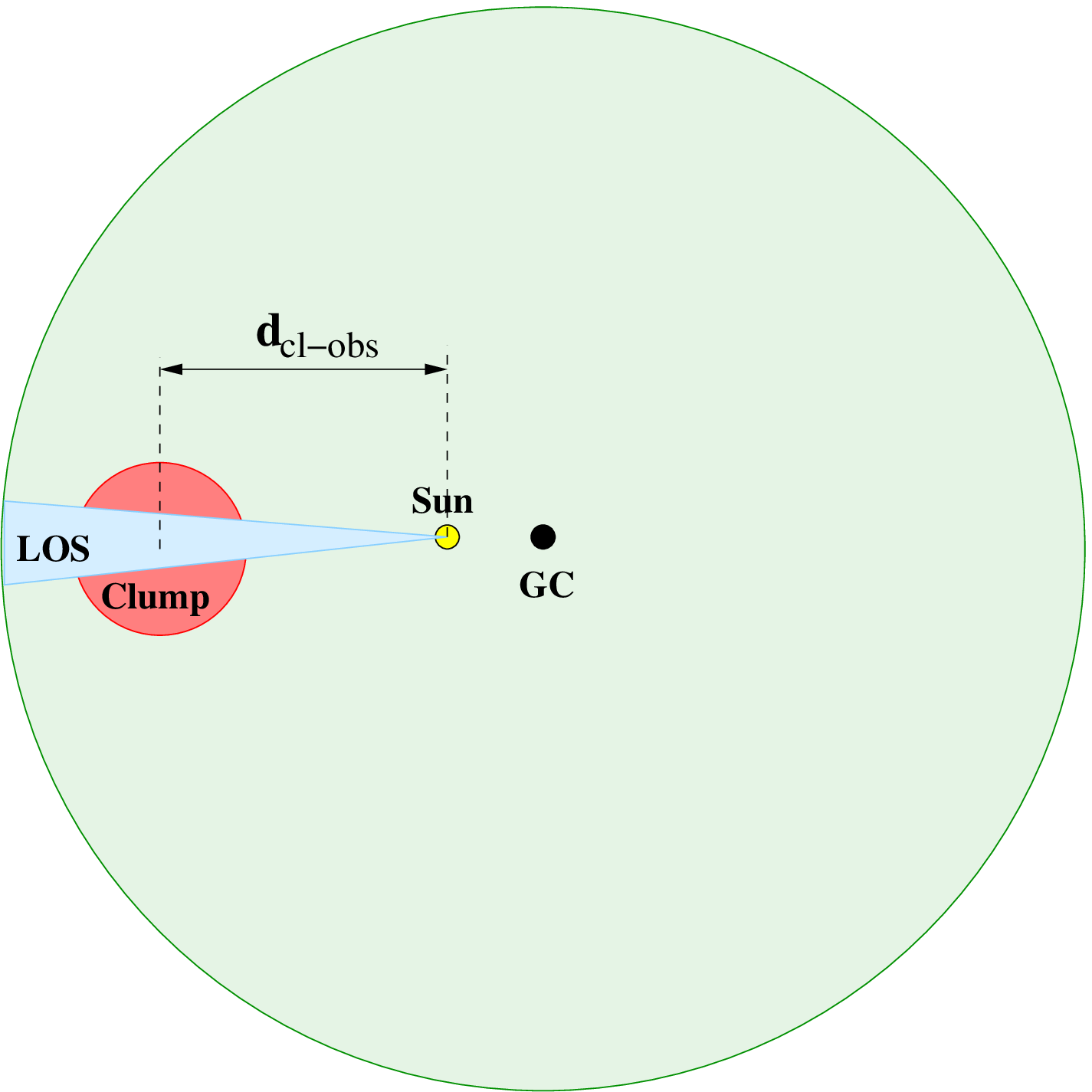}
\hspace{0.14\textwidth}
\includegraphics[width=0.48\textwidth]{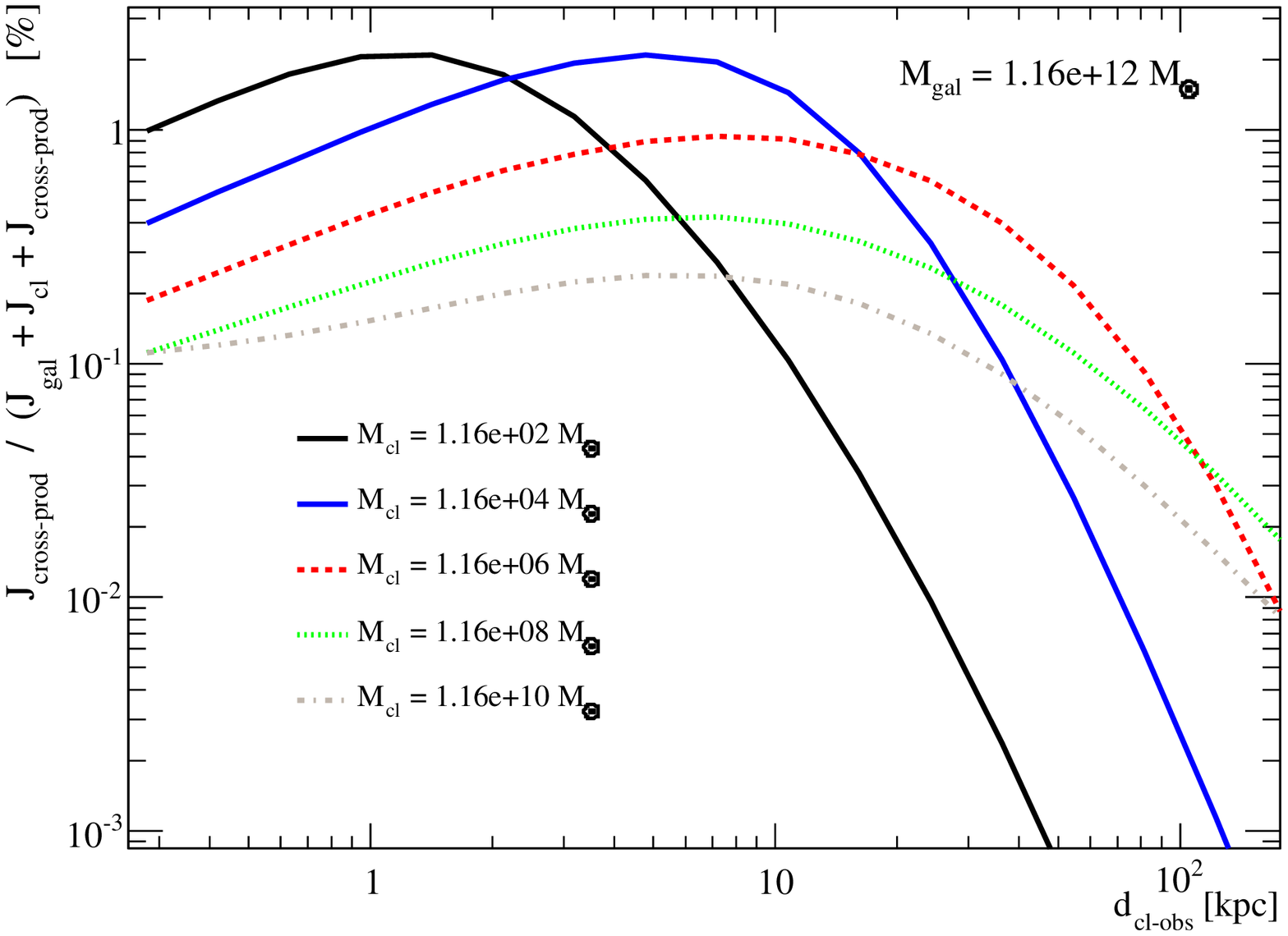}
\includegraphics[width=0.37\textwidth]{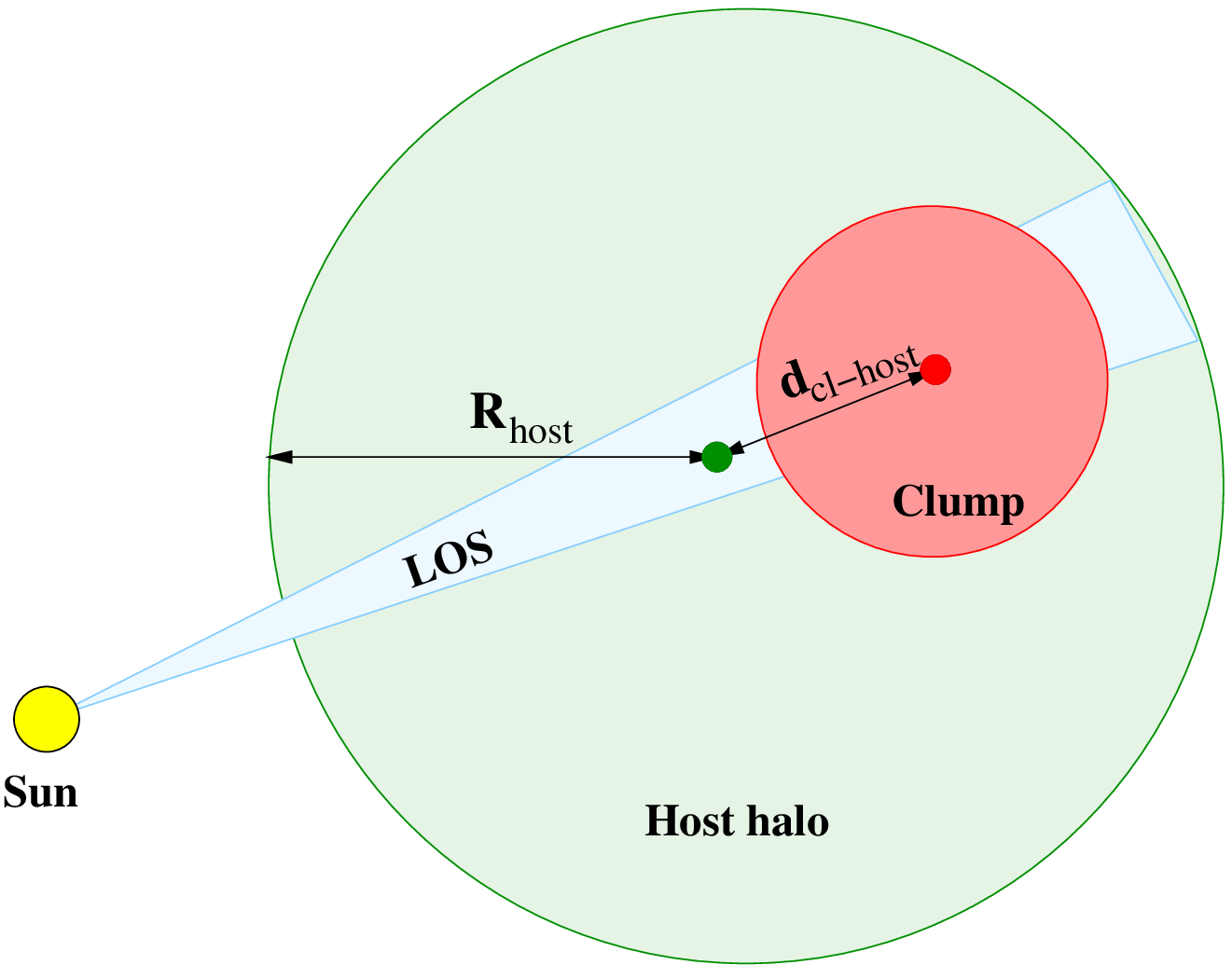}
\hspace{0.08\textwidth}
\includegraphics[width=0.48\textwidth]{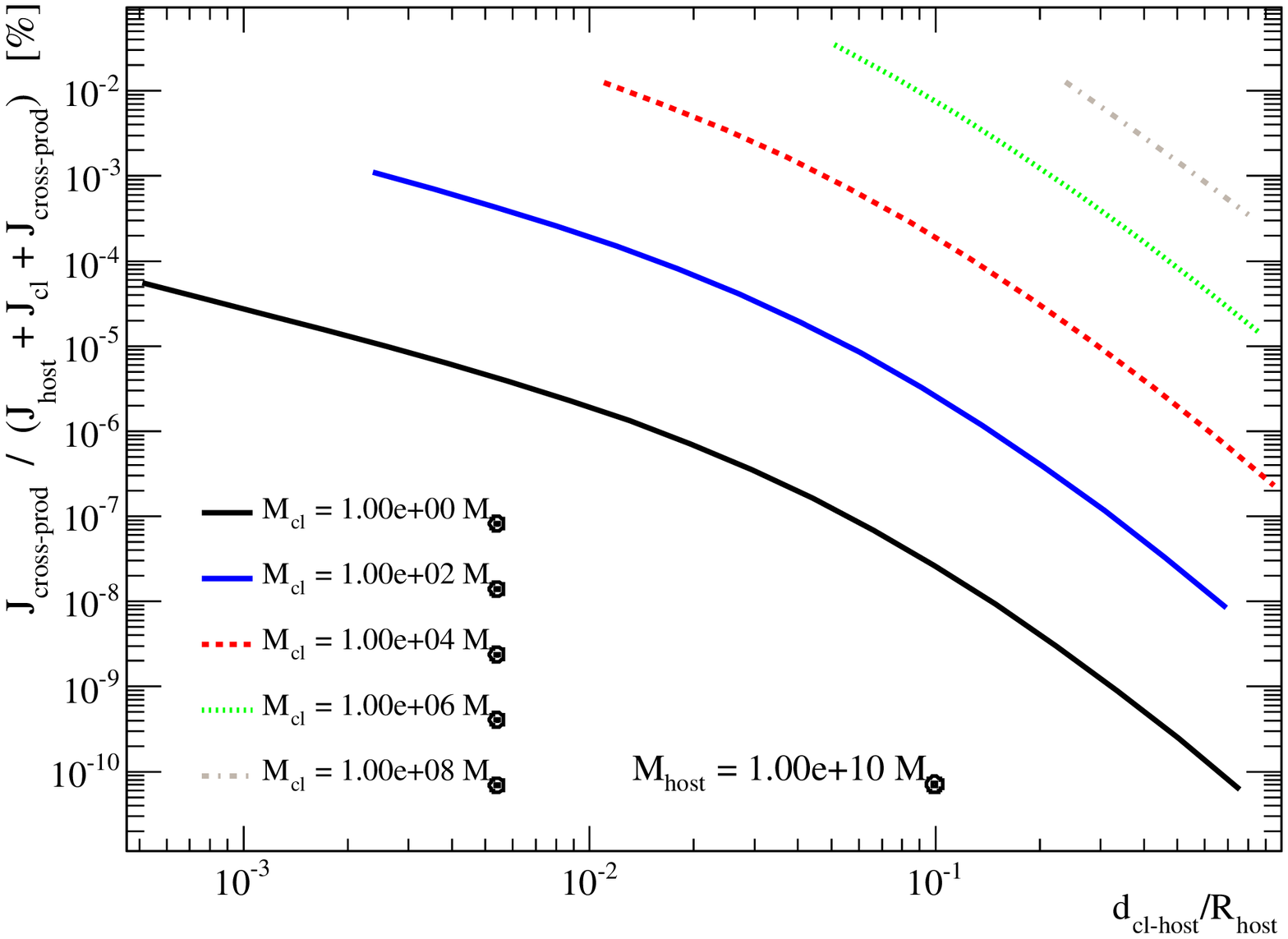}
\caption{Relative importance of the cross-product w.r.t to the total J
  calculation for i)  a clump embedded in the Galactic halo
in the anti-centre direction (top panels) and ii) a sub-clump in DM halo
located 100~kpc away (bottom panels). Several masses for the clump have been
considered and the integration angle is $\alpha_{\rm int}=0.1^\circ$.}
\label{fig:error_crossprod}
\vspace{-0.5cm}
\end{figure*}
%
\section{Conclusions \label{sec:conclusion} }

The {\sc clumpy} code is optimised to calculate the line-of-sight integration
of the density and/or the density square of the DM distributions
(smooth and clump distributions). It can be used by (i) experimentalists
looking for realistic $\gamma$-ray skymaps to calculate their new instrument
sensitivity, or simply to use them in model/template analyses for the DM diffuse
emission near the Galactic centre (or for dSphs), (ii) for  astrophysicists working
on the reconstruction of the DM content of dSphs who wish to calculate the J factor
as a by-product of their study, (iii) theoreticians who want to plug their preferred
particle physics model and see what is the corresponding $\gamma$-ray flux in
the Galaxy/dSph. The code is fully documented and provides many functions
that makes it versatile and flexible.

Many improvements are possible for a future release. The simplest
and more straightforward ones to consider are:
\begin{itemize}
  \item Extra-galactic signal from the local structures (e.g. local clusters, Nezri et al. in prep.),
  but also from the high-redshift isotropic contribution. 
  \item The possibility to have the so-called Sommerfeld-like
  enhancements. This requires to couple the calculation of the
  spatial density of DM to the position-dependant velocity of the
  DM particles.
  \item Neutrino annihilation spectra and fluxes.
\end{itemize}
Some other improvements are intrinsically more difficult to add, as they
depend on astrophysical inputs that are not well determined.
Indeed, in the absence of a smoking gun signature for indirect detection 
(e.g. a line detection), multi-wavelength and multi-messenger analyses
seem to be a compulsory approach. To do so, one needs to take into account
\begin{itemize}
  \item the anti-proton and anti-deuteron signal;
  \item electron and positron production;
  \item secondary multi-wavelength emissions from leptons.
\end{itemize}

Finally, the {\sc clumpy} code could be coupled to some particle physics code to
get more specific annihilation spectra  (e.g., darkSUSY, Gondolo et al.
\citep{2004JCAP...07..008G}; MicrOMEGAs, Belanger et al. \citep{2010arXiv1004.1092B}), or to
some cosmic-ray propagation code (e.g.
GALPROP\footnote{http://galprop.stanford.edu/}, 
DRAGON\footnote{http://www.desy.de/~maccione/DRAGON/} or USINE, Maurin et al., in
preparation) to compare with the Galactic astrophysical backgrounds. This will be
considered for a second release of the code.

\section*{Acknowledgements}
We thank E. Nezri and J.-C. Lambert for their useful comments on the code. We also thank the anonymous referee the many 
constructive comments that improved both the code and the paper.

 \appendix
\section{Geometry and change of coordinates \label{app:geom}}

We define two frames of reference: one attached to the observer the origin of
which is the Earth, and the other linked to the Galactic Centre  (see
Fig. \ref{fig:geom}). The frame related to the Earth $\left( \rm{E}, l,
\psi, \theta \right)$ is spherical, and is identified to the  Galactic
coordinate system. $l$ is the distance to the Earth, the Galactic
longitude $\psi$ is measured in the plane of the Galaxy using  an axis
pointing from the Earth to the Galactic Centre, the Galactic latitude is
measured from the plane of the Galaxy to the object. 

The Galactic Centre frame $\left( \rm{GC}, x, y, z\right)$ is a cartesian
frame, where the $\left( \rm{GC}, x\right)$ and  $\left( \rm{GC},
y\right)$ axes are in the Galactic plane, and the $\left( \rm{GC},
z\right)$ axis is pointing to the North Galactic pole.  In this frame, the
position of the Earth is $\left(-\Rsol, 0,0\right)$, where $\Rsol$ is the
distance between the Earth and the Galactic Centre.

The change of coordinates formulae are given in the documentation
({\tt geometry.h}).

\section{Cross-product of individual clumps \label{app:crossprod}}
In \S\ref{subsec:calc_J} it has been shown that the cross-product between the
underlying smooth DM density and that of all the clumps in the line of sight (when
using an average description) can significantly contribute to the value of J in
the line of sight. It is interesting to check what contribution it brings when
considering one clump only (rather that the entire distribution). We use the same
equations as in \S\ref{subsec:calc_J} but written for one clump only and
numerically integrates J. Two configurations are explored: i) a clump embedded in
the Galactic halo (fig.~\ref{fig:error_crossprod}, top panels) and ii) a clump
embedded in a host halo located away from the sun (fig.~\ref{fig:error_crossprod},
bottom panels).

In fig.~\ref{fig:error_crossprod}, the relative error made when
neglecting the cross-product is plotted against i) the distance to the
clump for the first configuration (top), ii) the distance of the clump
to the centre of its host halo (bottom). In both cases, these curves
show that, as far as one individual clump is concerned, the cross-product is completely
negligible, whatever situation occurs: it amounts at most to $\sim 1\%$.

\bibliographystyle{cpc}

\end{document}